\documentclass{IEEETran}
\usepackage[noadjust]{cite}
\usepackage{amsmath,amssymb,amsfonts}
\usepackage{algorithmic}
\usepackage{graphicx}
\usepackage{textcomp}
\usepackage[table]{xcolor}
\usepackage{xcolor, soul}
\def\BibTeX{{\rm B\kern-.05em{\sc i\kern-.025em b}\kern-.08em
    T\kern-.1667em\lower.7ex\hbox{E}\kern-.125emX}}
\begin{document}
\title{Class-E, Active Electrically-Small Antenna for High-Power Wideband Transmission at the High-Frequency (HF) Band}
\author{Nathan Strachen, John Booske, \IEEEmembership{Fellow, IEEE}, and Nader Behdad, \IEEEmembership{Fellow, IEEE}
\thanks{This material is based upon work supported by the Office of Naval Research
under ONR Awards No. N00014-16-1-2098. N. Strachen, J. Booske, and N. Behdad are with the ECE Department of the University of Wisconsin Madison.
(email: Strachen@wisc.edu; jhbooske@wisc.edu; behdad@wisc.edu).

}}
\maketitle

\begin{abstract}
Antennas operating at the high-frequency (HF) band (3-30 MHz) are frequently electrically small due to the large wavelength of electromagnetic waves (10-100 m). However, the bandwidth-efficiency products of passively matched electrically small antennas (ESAs) are fundamentally limited. Wideband HF waveforms using bandwidths of 24 kHz or more have recently received significant attention in military communications applications. Efficiently radiating such signals from conventional passive ESAs is very challenging due to fundamental physical limits on bandwidth-efficiency products of ESAs. However, active antennas are not subject to the same constraints. In this work, we present the design and experimental characterization of a high-power, active ESA with enhanced bandwidth-efficiency product compared to {that of} passively matched ESAs. Specifically, the proposed active ESA can radiate wideband HF signals with banwidths of 24 kHz or more, with total efficiencies up to 80$\%$, and radiated power levels approaching 100 W. Our approach uses a highly-efficient, integrated class-E switching circuit specifically designed to drive an electrically small, high-Q HF antenna over a bandwidth exceeding 24 kHz. Using a high-Q RLC antenna model, we have successfully demonstrated wideband binary ASK, PSK, and FSK modulations with the proposed class-E switching architecture. Experimental results indicate that the bandwidth-efficiency product of this class-E active antenna is 5.4-9.8 dB higher than that of an equivalent passive design with the same data rate, and bit-error-rate (BER).
\end{abstract}

\begin{IEEEkeywords}
Transmitting antennas, electrically small antennas, switching amplifiers, active antennas.
\end{IEEEkeywords}

\maketitle

\section{Introduction}
\label{sec:introduction}
\IEEEPARstart{T}{he}
high frequency (HF) band in the range of 3-30 MHz is used in many military communications due to its unique long-distance propagation characteristics \cite{HFprop}. In recent years, there has been a growing interest in wideband HF communications for beyond-line-of-sight communications applications \cite{WBHF_art}-\cite{RFPerformanceImplicationsofWidebandHFWaveforms}. Wideband HF communication protocols using bandwidths of 24 kHz - 48 kHz have been implemented \cite{AnExt_WB_HFCapabilities}. However, practical implementation of wideband HF communications is not without challenges. One significant challenge stems from the fact many HF antennas used in mobile systems must have relatively  small physical dimensions. Due to the long wavelengths of the electromagnetic (EM) waves at these frequencies (10-100 m), such antennas are often extremely electrically small. Over the years, many studies have examined the fundamental limitations on the performances of electrically-small antennas (ESAs) \cite{b1}\nocite{Wheeler}\nocite{McLean}--\cite{Lopez}. These studies have demonstrated that there is a lower bound for the quality factor (Q) of an antenna with a given electrical size. Therefore, ESAs generally have input impedances with very low radiation resistance, and very high reactance values \cite{Lopez}. Significant losses can occur when trying to impedance match a high-Q transmit antenna with a passive matching network. Because the radiation resistance is so small, even a small amount of loss resistance can significantly degrade the gain of the ESA \cite{Lopez}, \cite{Hansen}. Furthermore, passive matching networks can introduce significant losses when used to match a high-Q, high-VSWR ESA. Due to the strong mismatch between input and output VSWRs {of an electrically-small antenna}, the losses present in the matching network can be significantly magnified \cite{Hansen}-\cite{Tingyen}. {Since both gain and the effective isotropic radiated power (EIRP) of the antenna are directly proportional to its total efficiency, any reduction in total efficiency will proportionally reduce antenna gain and EIRP {\cite{balanis_2016}}}. Due to all of these efficiency-degrading effects, commercially available HF transmitting antennas have typical reported gain values of $-$20 dBi at the lower end of the HF band. For example, a commercially available 5.07-m long HF whip antenna with a matching unit, is reported to have a gain of $-$20 dBi at 5 MHz, and $-$25 dBi at 3 MHz (Trival antene, AD-4/WB-MHD-4.8) \cite{trivalantene}. Due to such low  efficiencies, very large RF input power levels are required to radiate moderate RF power levels from the antenna. For example, to achieve an EIRP of 1 W from the aforementioned commercially-available antenna \cite{trivalantene} at 3 MHz, an RF input power level of 316 W is needed. Such large RF sources may not be practical, especially for mobile applications where physical space, battery capacity, and generator fuel are limited. However, even if the effects of loss could be mitigated, a passively-matched HF antenna will then have a limited bandwidth due to the very high quality factor of the antenna. The Bode-Fano limit constrains the achievable bandwidth of a passive, lossless, reciprocal matching network for an antenna with a given quality factor \cite{b2}-\cite{Youla}.

To surpass these fundamental limits, a significant amount of research has been done in the area of active, non-LTI (linear time invariant) antennas to enhance the bandwidth-efficiency products of ESAs. Two prominent techniques in these area include the use of non-Foster active impedance matching and direct antenna modulation (DAM) techniques. Non-Foster impedance matching involves synthesizing a negative inductance or capacitance from active components to cancel the reactance of an ESA over a large bandwidth \cite{Sussman1}. Significant improvements in bandwidth and gain have been experimentally demonstrated in the literature \cite{Tingyen}-\cite{Sussman2}. In \cite{Sussman2}, low-power class-A and class-B negative impedance converters (NICs) were designed for ultra-wideband matching from 15-30 MHz for transmit applications. They demonstrated transducer gain improvements up to 22.5 dB over conventional passive matching, and a maximum radiated power level of 4 dBm at 23 MHz from a 0.61 meter monopole antenna. While these improvements in gain and bandwidth are impressive, non-Foster matching suffers from several serious limitations. The first, and most relevant limitation to HF operation, is their limited power handling capability \cite{Tyler1}-\cite{Tyler2}. Because of the high-quality factor of an ESA, large voltages and currents are required at the antenna terminals to radiate significant power. These voltages and currents must be sustained by the active devices that comprise the non-Foster impedance matching circuit, placing severe power limitations on the non-Foster matched active antenna.

In \cite{Sussman1}, the problem of high voltage swing was addressed by designing a resonant class-C NIC that significantly reduced the voltage levels on the active devices. They experimentally demonstrated operation from 21-22.2 MHz, with an average radiated power level of 1.3 W from a 2-ft monopole antenna \cite{Sussman1}. However, significant limitations with class-C non-Foster circuits still exist. In this architecture, the bandwidth of the antenna is limited by the voltage capability of the active devices \cite{Sussman1}. Driving the circuit too far from the resonant frequency of the antenna can cause the voltage to exceed the maximum rating of the active devices. Higher power levels exacerbate the problem. Furthermore, high-power levels can cause instability and nonlinear distortion in non-Foster impedance matching networks, thereby effectively limiting the EIRP of the transmitting antenna \cite{HPNIC}. Additionally, the positive feedback in non-Foster circuits makes them prone to instability and oscillation at different frequencies \cite{HPNIC}-\cite{SA}.

Time-varying matching networks and direct antenna modulation provide another approach for realizing wideband active ESAs \cite{DAM_trans_fSK}-\nocite{HSpulseRad}\nocite{FSKdam_2020}\cite{Transitent_state_ant}. Direct antenna modulation involves the use of strategically timed switches within the matching network to enable rapid digital modulation in frequency, phase or amplitude. Significant improvements in performance have been experimentally demonstrated \cite{Transitent_state_ant}-\cite{HEVHFESABFSKDAM}. In \cite{HEVHFESABFSKDAM}, the authors demonstrated the transmission of a wideband BFSK signal centered at 75 MHz, at bit rates of 700 kb/s–7 Mb/s using a DAM circuit with a capacitively-loaded loop antenna. They demonstrate a bandwidth-efficiency product that is 5$\times$ greater than what would be achieved if the loop antenna was passively matched. In \cite{Transitent_state_ant}, a capacitively loaded loop antenna with DAM capable of generating double-side-banded amplitude modulation was reported to have achieved significant bandwidth improvements over passive matching. However, implementation of DAM is not without its challenges. DAM systems are generally limited in their radiated power capability. The switches in a DAM system must be able to sustain the large voltage and current levels at the terminals of the ESA. Additionally, implementation of DAM requires very fast switches with little parasitic capacitance. These challenges have been outlined in \cite{DAMParstics}, showing that non-ideal switches can cause unwanted oscillations. Furthermore, most DAM ESAs reported in the literature have low radiated power levels, which is a limitation for HF applications in which radiated power levels ranging from tens to hundreds of Watts are generally required.

In this work, we report on the design of an active electrically-small antenna that offers two important advantages over competing techniques. Firstly, the active ESA presented in this work is capable of radiating high-power levels resulting in high EIRP values. This is particularly important for long-range HF communications applications. Secondly, it offers bandwidth-efficiency product improvement factors ranging from 5.4 to 9.8 dB over passively-matched ESAs, depending on the modulation type. The simultaneous achievement of these two attributes in a single active-ESA design is a key contribution of this work, which has not been reported elsewhere. Our approach is based on the concept of integrating an electrically-small antenna with a switch-mode power amplifier (PA). Commercially available power amplifiers are designed to work with a specific load impedance, most commonly 50 $\Omega$ and antennas must then be matched to 50 $\Omega$. In contrast, integrated antenna/amplifier design involves designing the amplifier and the antenna in tandem. The concept of designing antennas integrated with amplifiers has been examined in other contexts before for both transmitting {\cite{IntAmpD}}-{\cite{Class_AB_ESA}} and receiving {\cite{Superconduct}} antenna applications. In {\cite{Superconduct}}, the design of electrically-small receiving antennas integrated with superconducting detectors for broadband, receive-only applications is discussed. While the concept is interesting, the requirement of operation at liquid-helium-cooled temperature of $4.2^\circ$ K is a major barrier towards adaptation of this technology in most applications. In the area of active transmitting antennas, which is the exclusive focus of this paper, examining the literature reveals that very few previous works have focused on addressing the challenges of integrating power amplifiers with highly-reactive electrically small antennas. In \cite{Class_AB_ESA}, an electrically-small antenna integrated with a class-AB amplifier was reported. This ultra-wideband antenna was designed for transmission of signals with instantaneously wide bandwidths in the HF band. As a result, the active antenna was designed to operate in a linear class-AB mode. Due to the UWB operation requirement and the highly-reactive nature of the antenna, the average radiated power of this antenna was limited to only 108 mW. Additionally, while it provided an average $6.8\times$ improvement in bandwidth-efficiency product ($\beta\times\eta$) compared to a passively-matched antenna, its  bandwidth-efficiency product was still very limited ($\beta \times \eta=0.11\%$) due to its low efficiency \cite{Class_AB_ESA}. The low radiated power levels and extremely small efficiency values of this class-AB active ESA are a limiting factor in many HF applications that require high EIRP levels, particularly those operating on the lower part of the HF band (3-10 MHz).

Unlike previous work, the design approach presented in the present work allows for efficiently radiating very high power levels from an electrically-small antenna over bandwidths greater than 24 kHz. The proposed approach is based on integrating a switch-mode, class-E amplifier operating in its suboptimum mode of operation with a resonant electrically-small antenna. Unlike \cite{Class_AB_ESA}, in the present design, RF carrier generation and modulation are both performed at the active antenna stage. We have experimentally demonstrated the ability to generate binary amplitude shift keying (BASK), binary frequency shift keying (BFSK), and binary phase shift keying (BPSK) modulations. Moreover, we have experimentally demonstrated that the proposed design offers significant bandwidth-efficiency product\footnote{We define the efficiency as the radiated power divided by the total DC and gate driver input power. This is somewhat similar to power added efficiency in conventional power amplifiers but not identical since no RF input signal is delivered to the antenna and the modulation is performed at the antenna level.} improvements compared to conventional, passively-matched antennas. Experimental results demonstrate that the antenna can radiate high power levels (64 W or more) and BASK, BPSK, and BFSK waveforms with bandwidths greater than or equal to 24 kHz at 3 MHz and $\beta \times \eta$ factors that are 5.4 dB to 9.8 dB higher than their passively-matched counterparts (depending on the modulation technique) for the same bit rate, and bit-error-rate (BER). Moreover, the proposed design offers bandwidth-efficiency product and radiated power levels that are respectively $26\times$ and $593\times$ higher than those of the class-AB active ESA reported in \cite{Class_AB_ESA}.

\section{Design of an Active Transmitting ESA Using an Integrated Class-E Amplifier}
\label{sec:Concept}
\subsection{Principles of Operation for Optimum Class-E Amplifiers}
\begin{figure}[ht]
\includegraphics[width=8.5cm]{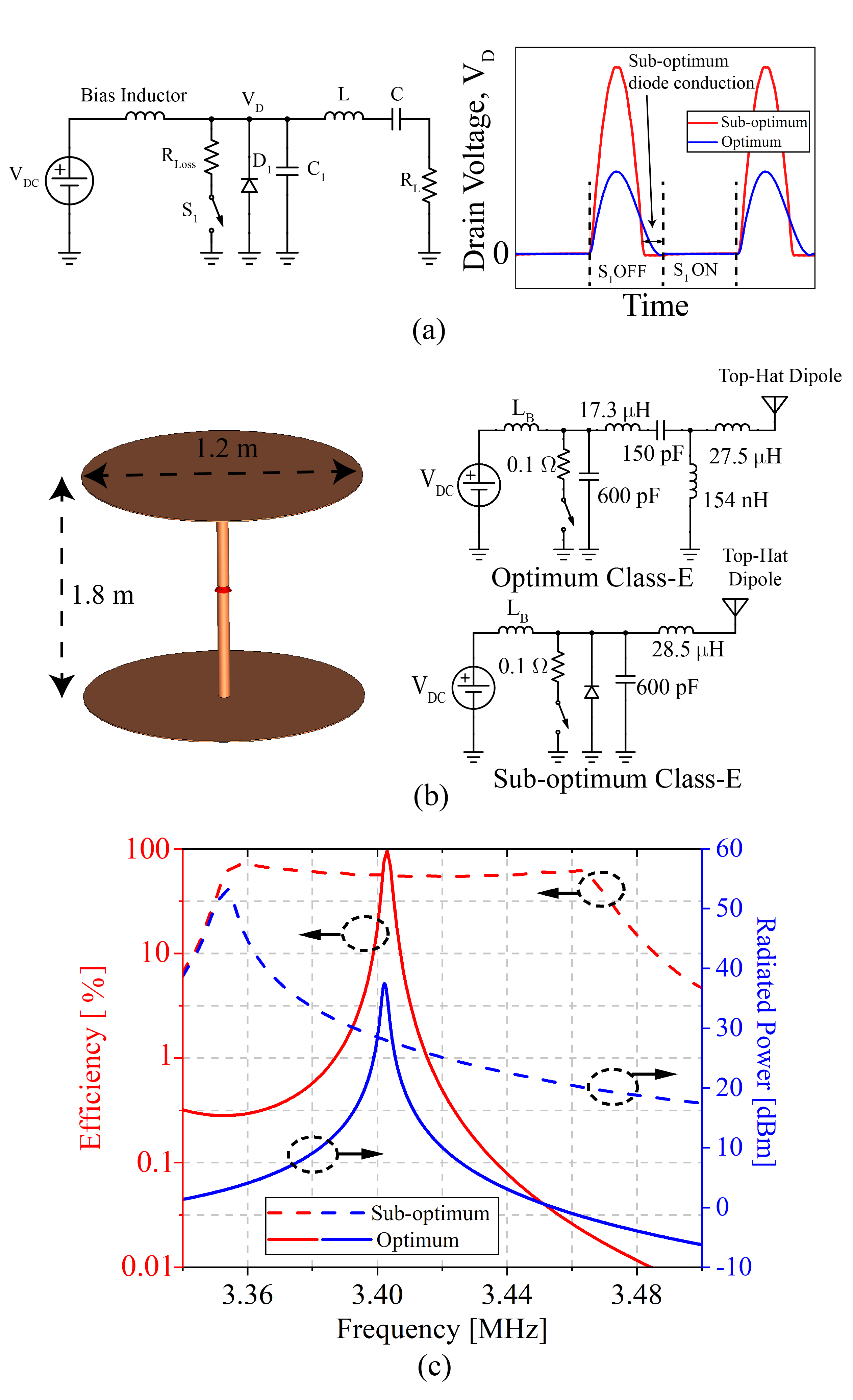}
\caption{(a) Class-E amplifier schematic and drain voltage. (b) {Integrated class-E amplifier with a top-hat-loaded electrically-small dipole antenna.} (c) Efficiency and radiated power for optimum and sub-optimum designs. Results show significant bandwidth improvement with sub-optimum design.}
\label{Concept}
\end{figure}

The class-E amplifier developed by Sokal \cite{Firstpap} and refined by Raab \cite{Raab} is a highly-efficient switch-mode amplifier with experimental efficiency values that can reach to 90$\%$ or more \cite{ClassEoverview}. The schematic of an ideal class-E amplifier is shown in Fig. \ref{Concept}(a). A typical class-E amplifier consists of an RLC load ($L$, $C$, and $R_{L}$ in Fig. \ref{Concept}(a)), a bias inductor, a switch (usually a MOSFET), and a parallel capacitance ($C_{1}$). Power MOSFETs usually contain an internal parallel diode ($D_{1}$), as shown in Fig. \ref{Concept}(a).

High efficiency can be achieved by carefully tuning the lumped element values to achieve zero-voltage-switching (ZVS), and zero-derivative-switching (ZDS) conditions. This is known as optimum class-E operation. The drain voltage across the transistor in a typical class-E amplifier with ZVS and ZDS conditions is shown in Fig. \ref{Concept}(a). The following relationship can be derived to determine the value of the optimum load resistance ($R_{opt}$) so that ZVS and ZDS switching conditions are achieved \cite{ClassEBook}:

\begin{equation}\label{OptimOp}
 R_{L}=R_{opt}= \dfrac{0.1836}{\omega C_{1}} = \dfrac{1}{5.447C C_{1} L \omega^{3}-1.525 C \omega}
\end{equation}
where $\omega = 2\pi f$, and $f$ is the switching frequency of the switch. To achieve ZVS and ZDS conditions, the frequency of operation must also be between the two resonant frequencies of the circuit, $f_{o1}$ (switch closed), and $f_{o2}$ (switch open):

\begin{equation}\label{fo1}
 f_{o1}=\dfrac{1}{2\pi \sqrt{LC}} < f < f_{o2}=\dfrac{1}{2\pi \sqrt{LC_{eq}}}
\end{equation}
where $C_{eq}$ is the equivalent series capacitance of $C_{1}$ and $C$:

\begin{equation}\label{Ceq}
 C_{eq}=\dfrac{C C_{1}}{C+C_{1}}
\end{equation}

\subsection{Integrated ESA Design with Optimum Class-E Amplifier}
\label{OptClassEDesign}
Ordinarily, optimum resistance values for ZVS and ZDS operation of class-E amplifiers are typically 10s of Ohms. In contrast, for our application, electrically small antennas present very low radiation resistance values (a few Ohms or less) \cite{Wheeler, Hansen}. Equating the optimum load resistance to a small radiation resistance requires $C_{1}$ to be very large (\ref{OptimOp}). This in turn would cause $f_{o1}$ and $f_{o2}$ to be very close, thereby severely limiting the bandwidth. Instead, to design an electrically small antenna with an optimum class-E amplifier, an impedance matching network must be used to transform the low radiation resistance of the antenna to a larger optimum resistance for the amplifier. However, given a lossless passive matching network and the high-Q antenna, the bandwidths of such matching networks are very small.

To demonstrate this, we designed and simulated an ideal (except for a 0.1 $\Omega$ switch loss resistance), optimum class-E amplifier with an electrically small, {top-hat-loaded dipole as shown in Fig. {\ref{Concept}(b).} We chose a top-hat-loaded dipole for this design example because it is a very good approximation of a Hertzian electric dipole with uniform current distribution along its length.  Such Hertzian dipoles are very well-understood and closed-form expressions for calculating their input impedance, efficiency, and radiated fields are available (e.g., see pp. 145-155 of {\cite{balanis_2016}}). The choice of this radiating element for this illustrative example does not limit the generality of the proposed design procedure in any way. In principle, the design process outlined in the remainder of the paper can be applied to any dipole or monopole type electrically small antenna.} The antenna is assumed to be made of copper and was simulated using full-wave electromagnetic simulations in CST Microwave studio to determine its input impedance and radiation resistance. We designed the class-E amplifier to operate near 3.5 MHz where the electrical length of the dipole antenna is $0.021\lambda_0$ given the physical length of 1.8 m. Such a small electrical length was chosen to illustrate the low bandwidth of ESAs. At 3.5 MHz, based on the CST simulated input impedance, the antenna is predicted to have an effective input capacitance of 79 pF and radiation resistance of 0.21 $\Omega$. The simulated radiation efficiency of the unmatched antenna was approximately 98$\%$, which indicates small conduction loss. {Real-world environmental factors, such as the presence of lossy earth close to the antenna, could decrease the radiation efficiency.} However, if true, this does not hinder the main point of this analysis, which is that given optimum class-E operation and a high-Q load, the bandwidth is very limited. Using (\ref{OptimOp}) and (\ref{fo1}), in addition to assuming a reasonable value of $C_{1}$=600 pF (based on typical parasitic capacitance values of HF power MOSFETs), the lumped element values necessary for optimum operation were determined. The load resistance required for optimum operation was calculated to be 50 $\Omega$, which is much larger than the radiation resistance of the antenna. This means that a matching network must be designed to convert the extremely low resistance of the antenna to 50 $\Omega$. To illustrate the bandwidth limitation of using a passive  matching network, we designed a matching circuit, shown in Fig. \ref{Concept}(b), using Optenni Lab software \cite{OptenniLab}. In Fig. \ref{Concept}(c), we plot the simulated radiated power and total efficiency ($e_{tot} = 10\log(P_{rad}/P_{DC})$) for the optimum design. As can be seen, the efficiency is very high (98$\%$) at approximately 3.4 MHz. However, the antenna's high quality factor results in a narrow, -3 dB transmitter bandwidth of only 1.8 kHz, which is far too low to support wideband HF communications.

\subsection{Sub-optimum Operation with Integrated Antenna Design for Wideband Performance}
For wideband operation with a high-Q antenna, optimum-operation of a class-E amplifier is not a viable approach as discussed in Section \ref{OptClassEDesign}. However, the class-E amplifier can be used in a different mode of operation that can significantly enhance the bandwidth of the antenna. This mode of operation is known as sub-optimum operation, and has been explored in several publications \cite{SubOpt_analysis}-\nocite{SubOpt_anal_nonlin}\cite{AnyDutyCycle}. Sub-optimum operation is achieved when only the ZVS condition is met. This yields an extra degree of freedom in the design, which allows for much higher operational bandwidths with high-Q antennas, but without sacrificing efficiency too much. Not having the ZDS switching condition only results in slightly more switching loss. Theoretical efficiencies approaching 95$\%$ are still possible in the sub-optimum mode \cite{Subopt_eff}. Unlike optimum operation, in the sub-optimum mode, the diode conducts for a brief period before the switch turns on \cite{ClassEBook}-\cite{SubOpt_analysis}. A typical drain voltage waveform across the transistor for sub-optimum operation is shown in Fig. \ref{Concept}(a). {The diode conducts in sub-optimum mode, because the drain voltage attempts to swing to a negative voltage, but is prohibited due to the conduction of the diode.} Sub-optimum operation occurs when $R_{L}<R_{opt}$. {In addition, based on analysis done in {\cite{SubOpt_analysis}}, the frequency of operation must satisfy ({\ref{SubOptfreq}}):}

 \begin{equation}\label{SubOptfreq}
 f_{o1}=\dfrac{1}{2\pi \sqrt{LC}} < f < f_{2}=-\dfrac{\sin(2 \phi)}{C_{1}\pi^3 R_{L}}
\end{equation}
{where $\phi$ is the phase of the current flowing through $R_{L}$. $\phi$ can be obtained by numerically solving:}
\begin{equation}\label{SubOptPhi}
 L\left( \dfrac{2\sin \left( 2\phi \right) }{C_{1}\pi ^{2}R_{L}}\right) ^{2}=-\dfrac{2\sin 2\phi }{C_{1}\pi ^{2}}\left( \cot \left( \phi \right) -\dfrac{\pi ^{2}}{4}\csc \left( 2\phi \right) \right) +\dfrac{1}{C}
\end{equation}
{Because the requirement for the load resistance in the sub-optimum mode of operation is that $R_{L}$ is less than the optimum resistance, there is no need for using an impedance matching network. This is in sharp contrast to the case for optimum operation where $R_L$ must assume a specific value as identified in {(\ref{OptimOp})}.} Instead, sub-optimum operation only requires adding a series inductor to the antenna {to achieve resonance}. This enhances the tunability of the transmitter across the HF band. In principle, any operating frequency can be selected based on tuning the series inductor, the shunt capacitor ($C_{1}$), and adjusting the digital gate switching frequency. We designed a sub-optimum class-E amplifier integrated with the top-hat dipole shown in Fig. \ref{Concept}(b). The simulation results for the circuit are shown in Fig. \ref{Concept}(c). As can be observed, the bandwidth of the sub-optimum amplifier is significantly greater than that of the optimum design. Specifically, the optimum design achieves an efficiency of 98$\%$ over a bandwidth of 1.8 kHz whereas the sub-optimum design achieves an efficiency of 60$\%$ over a bandwidth of $\sim$ 110 kHz. In other words, the sub-optimum class-E ESA design offers a factor of $~37.5\times$ improvement in bandwidth-efficiency product over the optimum design. The radiated power for the sub-optimum case reduces as frequency increases because the reactance of the RLC load increases {as we move away from the resonant frequency of the load}. This, however, is not a limiting factor as the radiated power can be adjusted dynamically by controlling $V_{DC}$ to maintain a constant radiated power level vs. frequency if desired. Doing this will not adversely impact the efficiency of the amplifier.

The efficiency of the sub-optimum class-E ESA, $e_{subopt}$, is primarily determined by the conduction loss resistance of the transistor, the loss resistance of the antenna, and the radiation resistance of the antenna:
\begin{equation}\label{EffSubopt}
e_{subopt}=\dfrac{P_{rad}}{P_{DC}+P_{SD}}\approx \dfrac{R_{rad}}{R_{loss}+R_{rad}+R_{la}}
\end{equation}
where $P_{rad}$ is the radiated power, $P_{DC}$ is the DC input power from $V_{DC}$, $P_{SD}$ is the power required by the switch driver circuitry to continuously switch the transistor, $R_{rad}$ is the radiation resistance, $R_{la}$ is the loss resistance of the antenna, and $R_{loss}$ is the conduction loss of the transistor switch. The power required to drive the gate of a power MOSFET ($P_{SD}$) is usually on the order of a few watts, and at higher radiated power levels $P_{SD}$ becomes a negligible contributing factor to the total efficiency. $P_{SD}$ can be estimated using analysis presented in \cite{ClassEBook}.

Based on (\ref{EffSubopt}), the class-E integrated amplifier operated in sub-optimum mode can achieve relatively high efficiency provided that the radiation resistance is larger than the loss resistances in the transistor or antenna. By carefully selecting the proper transistor with low drain-to-source ON resistance, and manufacturing the antenna from low-loss materials, high efficiency can be obtained without sacrificing the bandwidth of the antenna.

\section{Design of A Class-E, Active Electrically Small Antenna Prototype}
To validate our concept presented in Section \ref{sec:Concept}, we designed and built a sub-optimum class-E amplifier, integrated with a high-Q RLC load representing an electrically-small {resonant} antenna model. This RLC circuit is representative of any dipole- or monopole-type electrically-small antenna that is either self-resonant near its first resonance (e.g., a meandered electrically small dipole) or is made resonant by adding a series inductor to it (e.g., the top-hat loaded dipole shown in Fig. {\ref{Concept}(b)} with a series inductor added to it). For ease of prototyping and experimental evaluation, we built a high-Q lumped element RLC antenna model that mimics the frequency-dependent input impedance of an electrically small antenna. Using this dummy load as the antenna model allows for characterizing the performance of the proposed active antenna at high-power levels with simple benchtop experiments and eliminates the need for getting an FCC license for performing field tests at high-power levels.

\subsection{Lumped Element Antenna Model Design}
We used high-Q, high-voltage ceramic capacitors (AVX HQCE, 100 pF, voltage rating: 7.2 kV) to mimic the capacitance of an electrically small dipole type antenna similar to that shown in Fig. \ref{Concept}(b). By combining four capacitors in series and parallel, we synthesized a capacitance of approximately 100 pF, with a maximum voltage rating of 14.2 kV and a peak current rating of 14 A. This is reasonably close to the 79 pF capacitance that we calculated for the top-loaded dipole antenna shown in Fig. \ref{Concept}(b). The prototype was designed to work at 3 MHz, since the lower end of the HF band is where the problem of bandwidth limitations is most severe. We designed a large, high-Q air-core inductor of approximately 28 $\mu H$ to resonate with the high-Q capacitor at approximately 3 MHz. The inductor was constructed from 12 turns of 1 inch wide by 1.4 mil thick copper tape wound around a large five-gallon paint bucket, as shown in Fig. \ref{setup}. This load was connected to the class-E circuit through a 12 inch long, 16 AWG, 450 $\Omega$ ladder cable with SMA connectors soldered at the end.

\begin{figure}[ht]
\includegraphics[width=8.5cm]{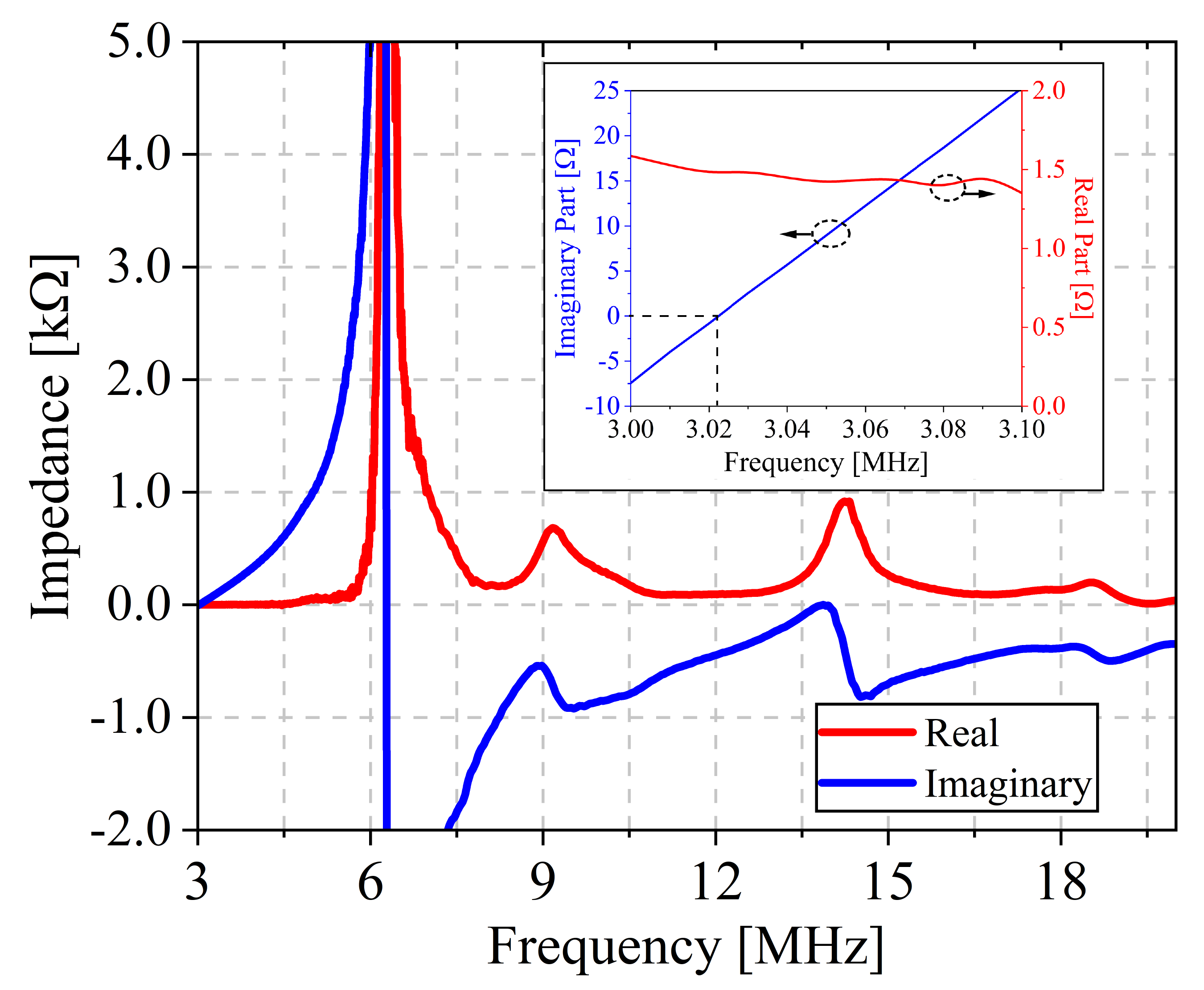}
\caption{Measured input impedance of the antenna model. The dashed lines show the location of the series resonance, at which point the imaginary part of the impedance is zero.}
\label{AntImp}
\end{figure}

The input impedance of the RLC antenna model (with the ladder cable included) was measured with a vector network analyzer (VNA, Copper Mountain Planar TR1300/1) and is shown in Fig. \ref{AntImp}. The inductor is in series resonance with the capacitor at approximately 3.02 MHz, and the resistance is on average 1.43 $\Omega$ near 3 MHz, which is representative of a small radiation resistance. The antenna model is representative of a high-Q ESA with poor bandwidth performance. Specifically, the maximum VSWR$\le$2 bandwidth that can be achieved given a lossless passive matching network is 6 kHz. At frequencies above 3 MHz, the reactance of the antenna model becomes much larger, reaching values on the order of several k$\Omega$, as shown in Fig. \ref{AntImp}. At just beyond 6 MHz, there exists a parallel resonance due to the parasitic capacitance of the ladder cable, and another resonance near 14 MHz due to the parasitic capacitance between the inductor windings. The magnitude of the impedance is large outside the band of interest, which minimizes the radiated harmonics.

\subsection{Active ESA Design and Measurement}
After constructing our high-Q antenna model, we designed a class-E amplifier specifically to be integrated with this load. A vertically-diffused MOSFET (VDMOS) DRF1201 transistor from Microsemi was chosen for the switching device. This device is specifically designed for high-power RF switching applications up to 30 MHz. It has a breakdown voltage of 1000 V, a maximum current rating of 26 A, and a low drain-to-source on resistance of $\leq$ 0.55 $\Omega$. It also has built-in gate driver circuitry, which enables switching the transistor on and off with a 5 V digital clock signal. The gate driver is powered by a 12 V power supply that consumes approximately 3.8 W at 3 MHz. The DRF1201 has a nonlinear internal drain-to-source capacitance that is dependent on the drain-to-source voltage of the transistor. Based on the datasheet, when the voltage is on the order of a hundred volts, the capacitance ($C_{1}$ in Fig. \ref{Concept}(a)) is approximately 250 pF. Using the values of the RLC antenna model and the equations presented in \cite{SubOpt_analysis}, $R_{opt}$ is determined to be in the range of 10s of Ohms and the frequency range over which suboptimum operation can be achieved is 3.024-3.31 MHz.

The last component in our design is the bias inductor. The inductor should have a sufficiently high impedance such that the AC current flowing through the inductor is relatively small compared to the DC current. However, larger inductors are generally more bulky and lossy. To evaluate a suitable realistic inductor, we measured the S-parameters of a commercially available, 25 $\mu$H toroidal inductor (API Delevan, PTHF-H). We then simulated this inductor in our design to determine its performance. Sub-optimum operation was observed in the simulation, confirming that the inductor provided sufficiently high impedance for proper operation of the circuit. This was verified by simulating the transistor drain-to-source voltage, and ensuring that the ZVS condition was met. The simulated DC current was 4.7 times as large as the peak-to-peak AC current ripple in the inductor.

\begin{figure}[ht]
\includegraphics[width=8.5cm]{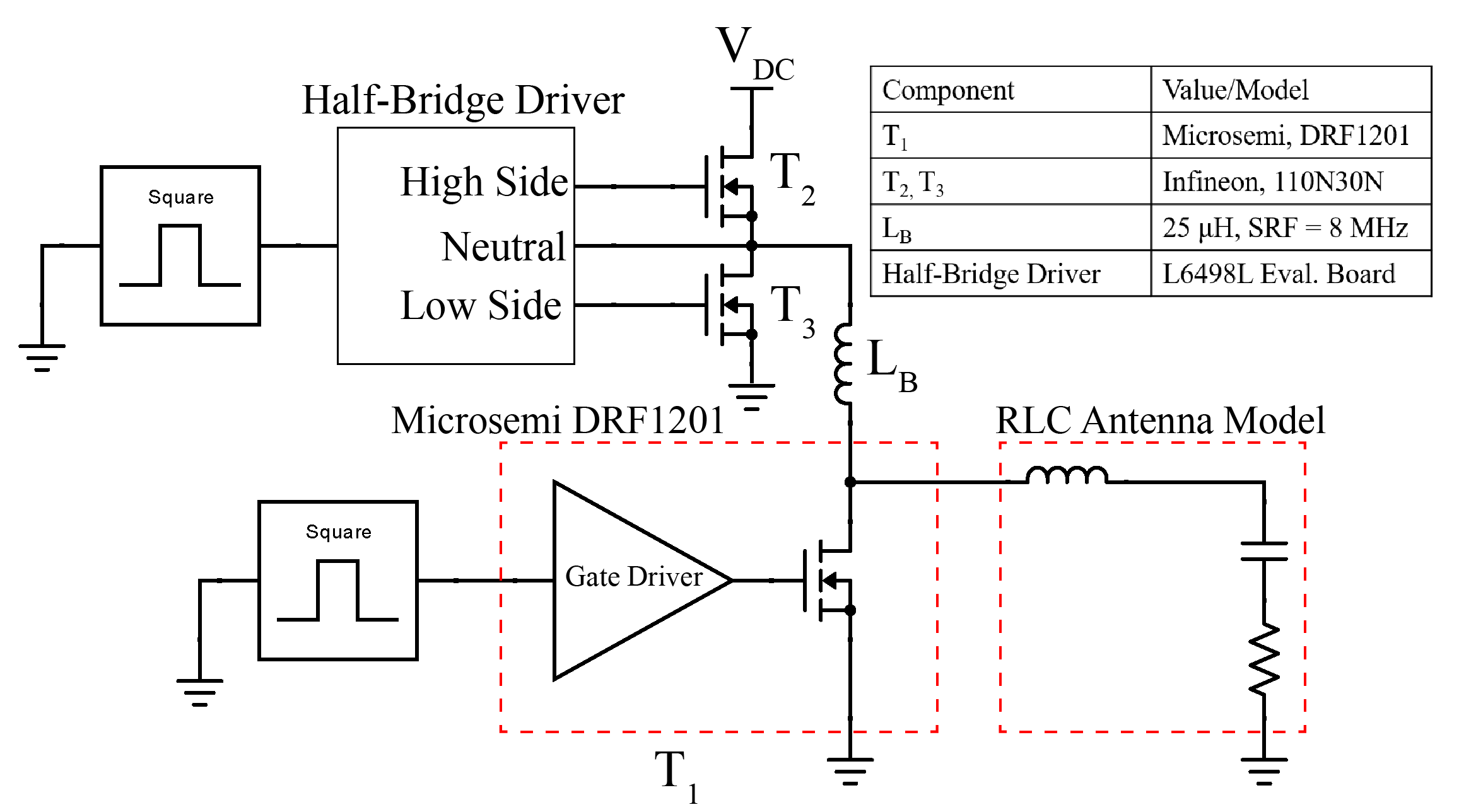}
\caption{Circuit schematic of the integrated sub-optimum class-E active antenna.}
\label{Schem}
\end{figure}

The complete design of our circuit is shown in Fig. \ref{Schem}. To demonstrate ASK modulation, the bias inductor, $L_{B}$, was connected to transistors $T_{2}$ and $T_{3}$. These transistors were used to turn the DC supply voltage on and off. If $T_{2}$ is on, and $T_{3}$ is off, then $V_{DC}$ is being supplied to the class-E circuit. If $T_{2}$ is off, and $T_{3}$ is on, then the supply voltage is turned off. $T_{2}$ and $T_{3}$ were alternated on and off with a commercial half-bridge gate driver (L6498L evaluation  board). For PSK and FSK modulation, $T_{2}$ and $T_{3}$ were unnecessary, and $L_{B}$ was directly connected to $V_{DC}$.

{The circuit shown in Fig. {\ref{Schem}} can be thought of as an ideal voltage source (i.e., one with zero output resistance) connected to a resonant load. This voltage source provides a time-varying waveform similar to that shown in Fig. {\ref{Concept}(a)} at the terminals of the resonant RLC load. Since the load is resonant, the current flowing in it would be sinusoidal as the higher-order harmonics are filtered out. Thus, the fields radiated by the antenna, which are proportional to its current, will also be sinusoidal. This ideal voltage source description of the problem clearly illustrates that no impedance matching between the antenna and the source is needed. The maximum power radiated by the antenna is simply limited by the amount of current the ideal voltage source can supply the load without damaging the main transistor.}

\begin{figure}[ht]
\includegraphics[width=8.5cm]{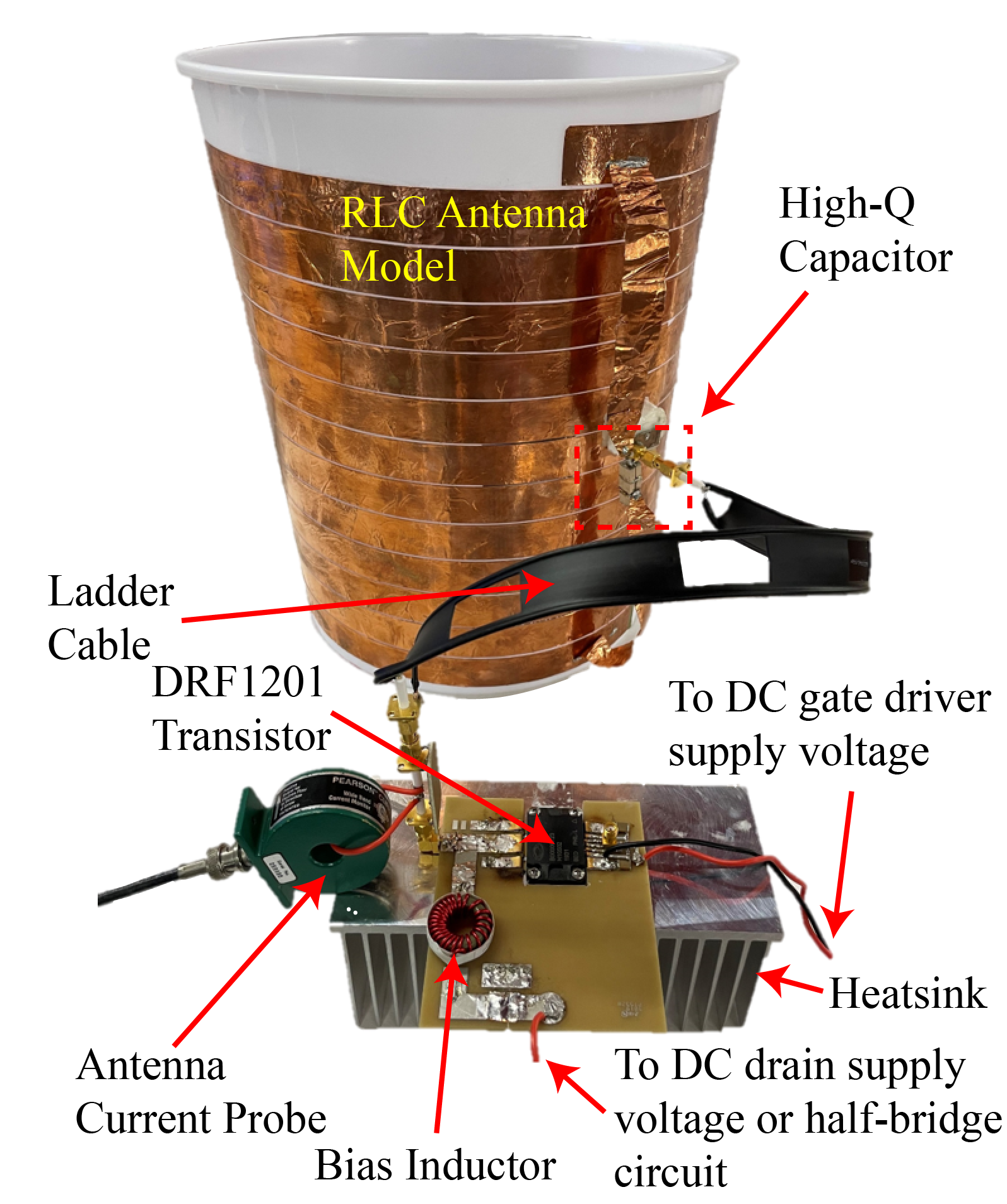}
\caption{Experimental setup for measuring the integrated sub-optimum class-E active antenna.}
\label{setup}
\end{figure}

\begin{figure}[ht]
\includegraphics[width=8.5cm]{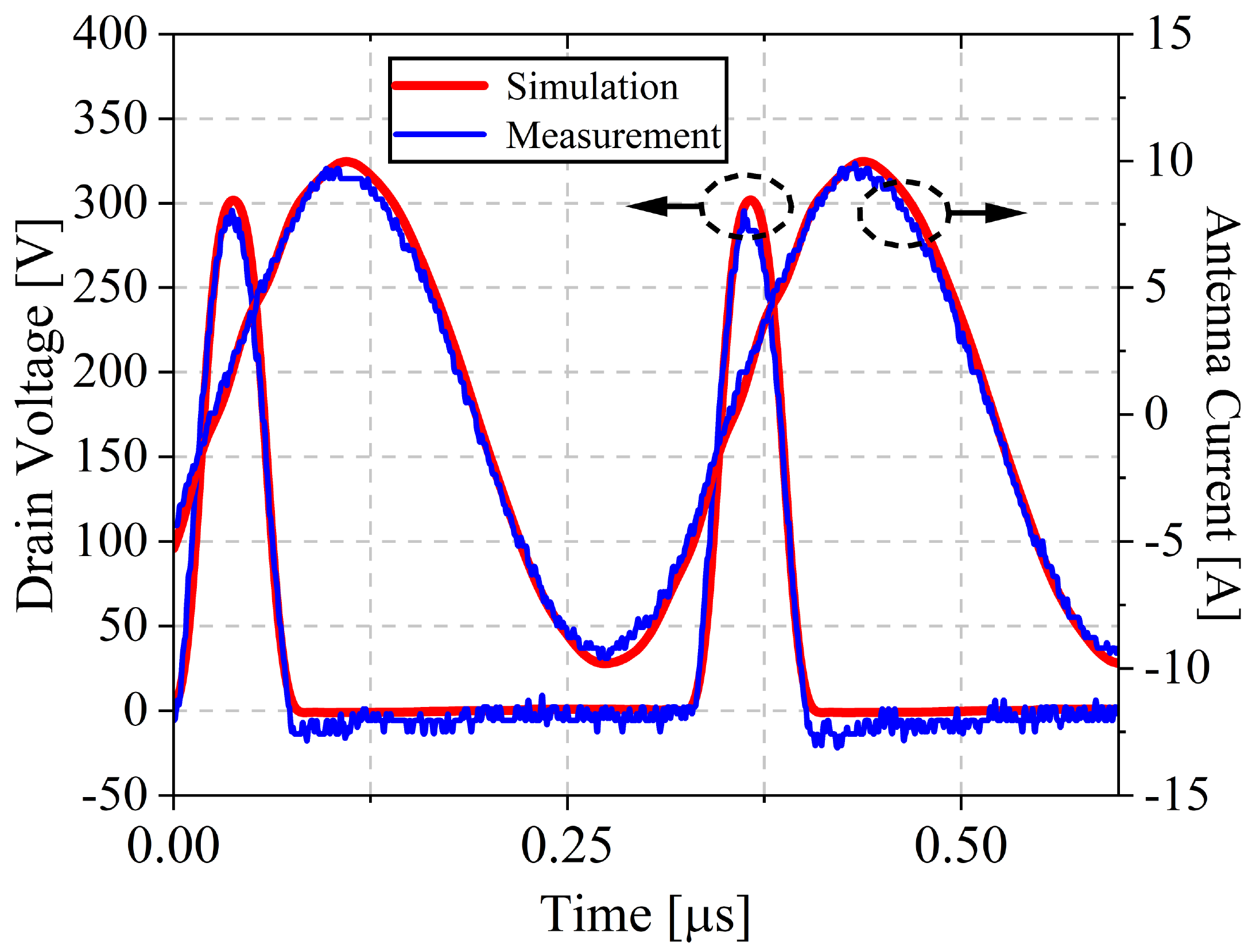}
\caption{Measured and simulated drain voltage and antenna current. $V_{DC}$ = 40 V, with a measured DC supply current of $I_{DC}$ = 1.96 A. The frequency of operation is at 3.045 MHz, and the total efficiency is 78$\%$, which includes DC driver power of 3.8 W.}
\label{CWop}
\end{figure}

We simulated our design with Keysight Advanced Design Systems (ADS). A nonlinear harmonic balance simulation was done to verify sub-optimum operation of the device. Realistic SPICE transistor models (provided by the manufacturer \cite{DRF1201}) and measured S-parameter models for the lumped elements were used in these simulations. Our simulation results for CW operation at a frequency of 3.045 MHz, and a DC supply voltage of 40 V are shown in Fig. \ref{CWop}. As can {be }seen from the drain voltage waveform, the ZVS condition was met, and therefore sub-optimum operation was achieved. The amount of power dissipated in the 1.43  $\Omega$ radiation resistance of the antenna model is 64 W, with peak drain-to-source transistor voltage levels of 300 V, which is within the transistor's maximum drain-to-source voltage rating of 1 kV. {The power transferred to the load can be increased by increasing the DC bias voltage of the transistor without exceeding its maximum rated drain-source voltage.}

After confirming the design through simulation, we constructed the complete prototype shown in Fig. \ref{setup}. We used a wideband current probe (Pearson, Model 411), and a high voltage oscilloscope probe to measure the antenna model's current and drain voltage respectively with an oscilloscope. The transistor was mounted on a large heatsink with thermal paste (Wakefield-Vette, 510-3M) with a heatsink-to-air thermal resistance of 0.56$^{\circ}$ C/W. The circuit board was made with FR4.

Our measurement results and comparison with simulation are shown in Fig. \ref{CWop}. As can be seen, excellent agreement between the measurement and simulation results is obtained. A large peak antenna current of 9.5 A was measured. Based on the measured peak antenna current, and the known antenna model capacitance value (C = 100 pF), the peak-to-peak voltage swing across the antenna model capacitor was approximately 10 kV ($2\times 9.5 \times 1/(\omega C) \approx 10^{4}$). This antenna voltage level was much larger than the measured peak transistor drain voltage of 300 V, as shown in Fig. \ref{CWop}. This is in contrast to many other active matching techniques, where the active devices must sustain all of the voltage across the antenna.

The measured total efficiency of the prototype is 78$\%$. The efficiency calculation includes the 3.8 W gate driver power, although it is relatively small compared to the 78 W drawn from the drain voltage supply. We verified the amount of power dissipated in the antenna load (the effective radiated power), through two independent methods to ensure the accuracy of our calculation. First, we computed the radiated power based on the known measured antenna current (9.5 A peak), and the effective radiation resistance of 1.43 $\Omega$ (measured with the VNA) of the antenna load ($P_{rad}=1/2\times9.5^{2}\times1.43 = 64.5$ W). Second, the radiated power was computed by multiplying the measured voltage and current in the time domain and averaging it over a cycle. This yielded a radiated power level of 63.4 W, which is very close to the level that we computed with the first method. The total efficiency of the amplifier is 78$\%$ using the first method, and 77$\%$ for the second. The strong agreement between these two methods increases confidence in the reported efficiency results.

We operated the transistor at a drain voltage value of 300 V, which is significantly lower than its maximum rating of 1 kV. This was done to avoid damaging the transistor or overheating the antenna load during our initial experiments. {Therefore, it is possible to operate the transistor at higher voltages and achieve even higher radiated power levels as indicated earlier.} Since radiated power is proportional to the square of $V_{DC}$, operating the transistor at its maximum drain voltage of 1 kV would result in a radiated power level of 700 W, and a peak antenna current of 31.7 A (average current of 20.2A). Since these values are within the rated operating range of DRF1201, the maximum achievable radiated power level and EIRP of this antenna are expected to be significantly higher than 64 W. {Unlike conventional antenna designs, maximizing the power delivered to the antenna examined in this design does not require any impedance matching and the conditions required for maximum power transfer are primarily determined by the properties of aforementioned ideal voltage source.} Nevertheless, even at 64 W, these radiated power levels achieved from this antenna are considerably higher than those of any other active electrically-small antenna reported in the literature.

The estimated 700 W {maximum }radiated power corresponds to the power dissipated in the 1.43 $\Omega$ radiation resistance of the antenna model. In a realistic antenna, some of this resistance will be loss resistance. However, as long as the loss resistance is small relative to the radiation resistance of the antenna, radiated power levels as high as 500-700 W are possible with the demonstrated active antenna architecture for DC input power levels of 700-900 W. In contrast, commercially available electrically-small HF antennas using passive matching circuits generally have gains of around -20 dBi at the lower end of the HF band \cite{trivalantene}. Thus, to radiate 500-700 W of power from such inefficient ESAs, commercially available power amplifiers capable of providing RF input power levels as high as 50-70 kW to the antenna would be required. {The low gains of conventional HF antennas (e.g., {\cite{trivalantene}}) are almost entirely attributable to the losses in their impedance matching networks used to match the highly-Q loads to 50 $\Omega$. The significant loss observed in the matching networks of these antennas is due to the fact that a typical ESA has a low input resistance and a high input reactance. To match this load to 50 $\Omega$ using discrete inductors and capacitors, very large circulating currents will exist in the circuit, magnifying the intrinsic loss of the reactive components to a much higher realized loss, thereby reducing the antenna's realized gain. This concept is referred to as matching circuit loss magnification (see pp. 39-58 of {\cite{Hansen}}).}

\begin{figure}[ht]
\includegraphics[width=8.5cm]{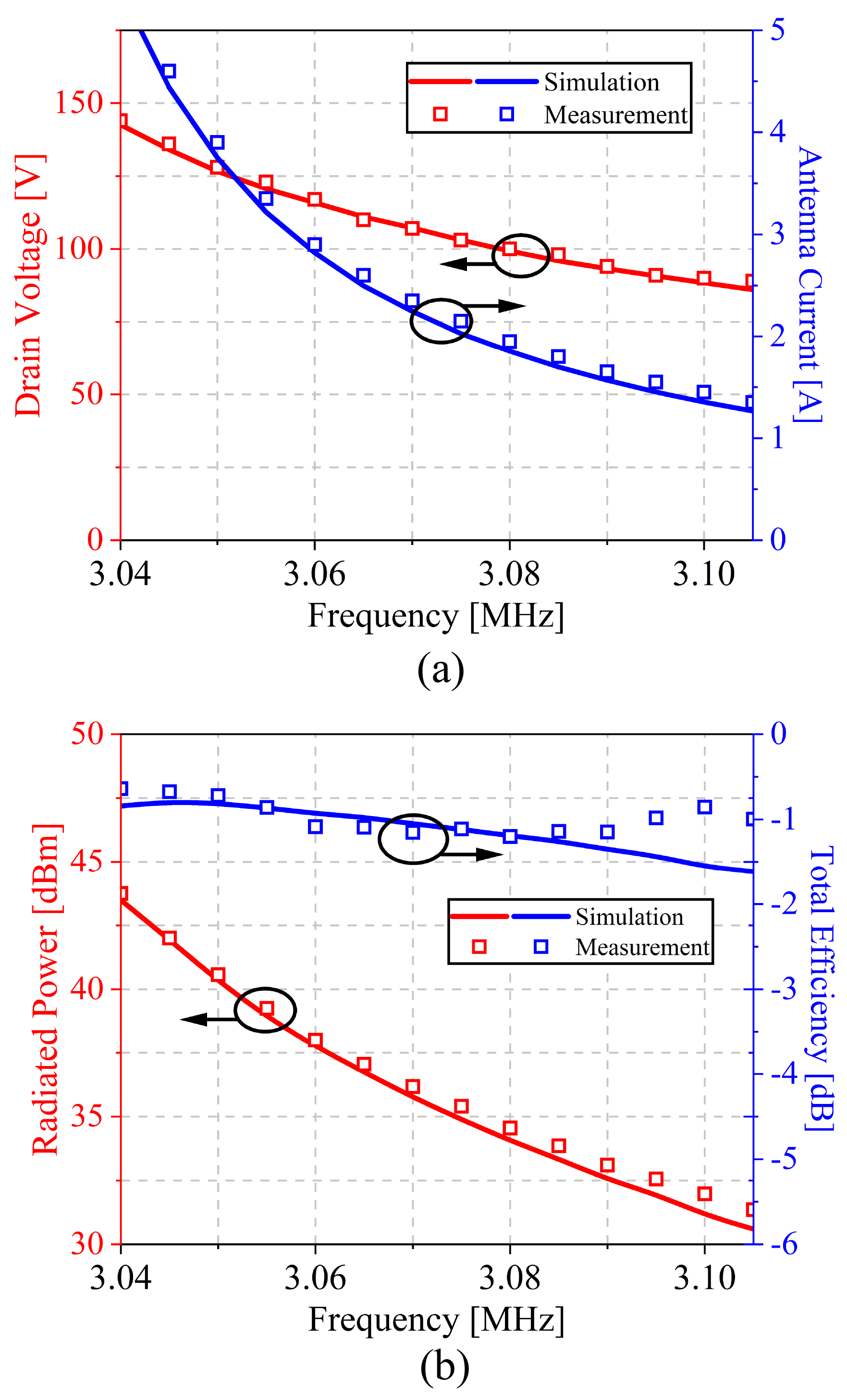}
\caption{Measurement and simulation results for CW operation across the operational bandwidth of the amplifier. $V_{DC}$ = 20 V for all measurements. The total efficiency includes DC driver power of 3.8 W.}
\label{CWop2}
\end{figure}

We also measured and simulated the antenna model at multiple frequencies to assess the accuracy of our simulations. The results of this are shown in Fig. \ref{CWop2}. A lower DC supply voltage of 20 V was used, for safety and ease of experimentation. Excellent agreement is shown between measurement and simulation. High efficiency is maintained over bandwidths greater than 60 kHz (see Fig. \ref{CWop2}(b)), which is sufficient for supporting wideband HF waveforms with instantaneous bandwidths of 24 kHz \cite{RFPerformanceImplicationsofWidebandHFWaveforms}. The bandwidth over which sub-optimum operation, and thereby high efficiency, is achieved is dependent on the radiated power level. At higher power levels, the voltage across the nonlinear drain capacitance is larger, which causes a decrease in the effective value of $C_{1}$, and increases the bandwidth. At a DC supply voltage of 20 V, the ZVS condition is no longer met at around 3.11 MHz, and the efficiency will begin to drop. However, larger supply voltages will allow for increasing the active ESA bandwidth. Fig. \ref{CWop2}(b) shows that for a constant DC supply voltage, the radiated power level decreases with a slope of approximately -0.15 dB/kHz due to the increase in the reactance of the RLC antenna load. This, however, is not a limiting factor since the radiated power level can be easily controlled by changing the amplifier's supply voltage ($V_{DC}$). Therefore, as long as the active ESA operates in its sub-optimum mode of operation and provides high efficiency, any change in the radiated power level versus frequency can be compensated for by dynamically changing $V_{DC}$ without impacting the efficiency. %The slope in the radiated power will affect the bandwidth and spectrum of a modulated signal as will be discussed in Section \ref{sec:Modulation}.

\section{Simulation-Based and Experimental Studies of Modulation Capabilities}
\label{sec:Modulation}
We have demonstrated ASK, PSK, and FSK modulation in both simulation and measurements. Time domain simulations in Keysight ADS were used to evaluate the modulation behavior and capabilities of the proposed active antenna. One of two switch models was employed in our simulations to model the behavior of the DRF1201 transistor used as as the switch. The first model was an ideal voltage controlled switch, a series loss resistance, a parallel diode, and parallel capacitance ($S_{1}$, $R_{Loss}$ = 0.55 $\Omega$, $D_{1}$, and $C_{1}$ = 250 pF in Fig. \ref{Concept}) that was used for simulating ASK and FSK modulation. Model parameters were tuned to provide the best fit of experimental data. However, this model did not provide as good an agreement with experimental data for PSK modulation. This is most likely due to the fact that the abrupt changes in phase that occur in PSK modulation cause  nonlinear effects in the transistor, which were not accounted for in the voltage controlled switch model. For this reason, a nonlinear SPICE transistor model provided by the manufacturer was used for PSK modulation, which yielded an improved agreement with our experimental results than the simple voltage controlled switch model.

\subsection{ASK Modulation}
The simplest form of modulation for a class-E amplifier is amplitude modulation through modulating the DC drain supply voltage. To implement ASK modulation, we used a half-bridge circuit, as shown in Fig. \ref{Schem}. The bandwidth that is achievable with ASK modulation is roughly dependent on the inductance of the bias inductor $L_{b}$, the DC input resistance of the amplifier, and the slope of the reactance versus frequency curve of the RLC antenna model near the operating frequency \cite{On_off_keyingpap}. The DC input resistance of the amplifier is equal to the DC input voltage, divided by the DC input current in CW operation ($R_{eff} = V_{DC}/I_{DC}$). The ratio of the bias inductance to the DC input resistance of the amplifier is a time constant that determines how quickly the drain voltage rises and falls. This rapidly rising and falling drain voltage is also filtered by the RLC antenna load. Because the reactance of the RLC load is higher at higher frequencies (see Fig. \ref{AntImp}), the upper sideband of the ASK spectrum will be more attenuated than the lower sideband.

\begin{figure}[ht]
\includegraphics[width=8.5cm]{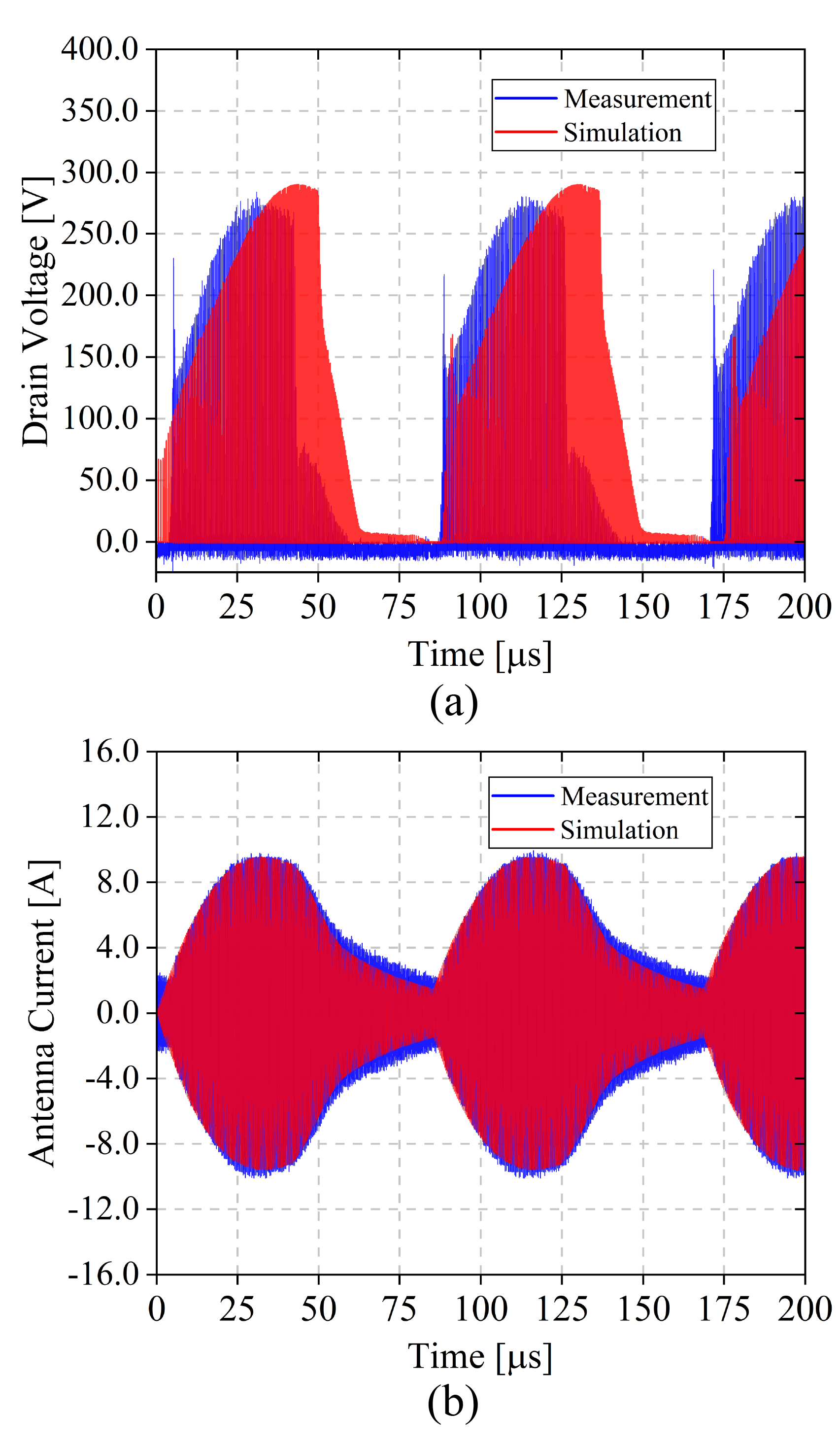}
\caption{Measurement and simulation results for ASK modulated signal with $f$ = 3.045 MHz. $V_{DC}$=35 V, and measured $I_{DC}$ = 1.14 A. The measured and simulated efficiency is approximately 40$\%$.}
\label{ASKmod}
\end{figure}

To test the wideband modulation behavior of the active antenna, we switched the drain voltage on and off at a rate of 12 kHz {(one period consisting of one on and one off cycle each with a duration of $\frac{1}{24}$ ms each) for a binary-ASK (BASK) signal with a bit rate of 24 kb/s}. Our experimental and simulation results are shown in Fig. \ref{ASKmod}. The magnitude of the drain voltage and antenna current correspond very well to our measurement results. The antenna current was simulated with a high degree of accuracy. The precise shape of the drain voltage waveform differs more significantly, likely due to the differences between the actual transistor and the transistor model used in our simulations.

In Section \ref{Nonlin_section} and \ref{passive_comp_sec}, we evaluate the radiated spectrum and discuss how it compares to conventional HF practice in passively matched antennas. ASK modulation is not commonly used for long-range HF communication since it is susceptible to noise and varying channel conditions. However, ASK modulation still may be useful in combination with FSK or PSK modulation for obtaining higher bit rates by synthesizing higher-order modulations that take advantage of the changes of both amplitude and phase or frequency (e.g., 16QAM).

Under ASK modulation, the measured and simulated 40$\%$ amplifier efficiency is much less than the 78$\%$ observed with CW operation. This results from the momentary loss of the ZVS condition when the drain voltage is rapidly rising and falling. When the drain voltage is low, the MOSFET's nonlinear capacitance is very high. Once the drain voltage builds up to a suitable level, the capacitance decreases and the ZVS condition is reached.

{We also performed a coherent demodulation of the simulated and measured antenna currents shown in Fig. {\ref{ASKmod}} to determine if the ASK modulation is performed correctly and estimate the expected error vector magnitude (EVM) of the modulated waveform at the transmitter. To do this, the time-domain waveforms of the simulated and measured antenna currents were exported to Matlab and a coherent ASK demodulation was performed. Following this process, we confirmed that the modulated waveform can be successfully demodulated and the transmitted bits recovered. The EVM was calculated from the simulation data since the response of the antenna could be easily simulated for a long pseudorandom bit sequence whereas the measurement was performed for a short periodic bit sequence (1010...) due to the limitations of the instruments available to us for performing the modulation in experiments. The simulated transmitter EVM was determined to be -24 dB at a bit rate of 24 kb/s without applying any equalization or additional post processing steps.}

\subsection{PSK Modulation}
\label{PSKmod_sec}
\begin{figure}[ht]
\includegraphics[width=8.5cm]{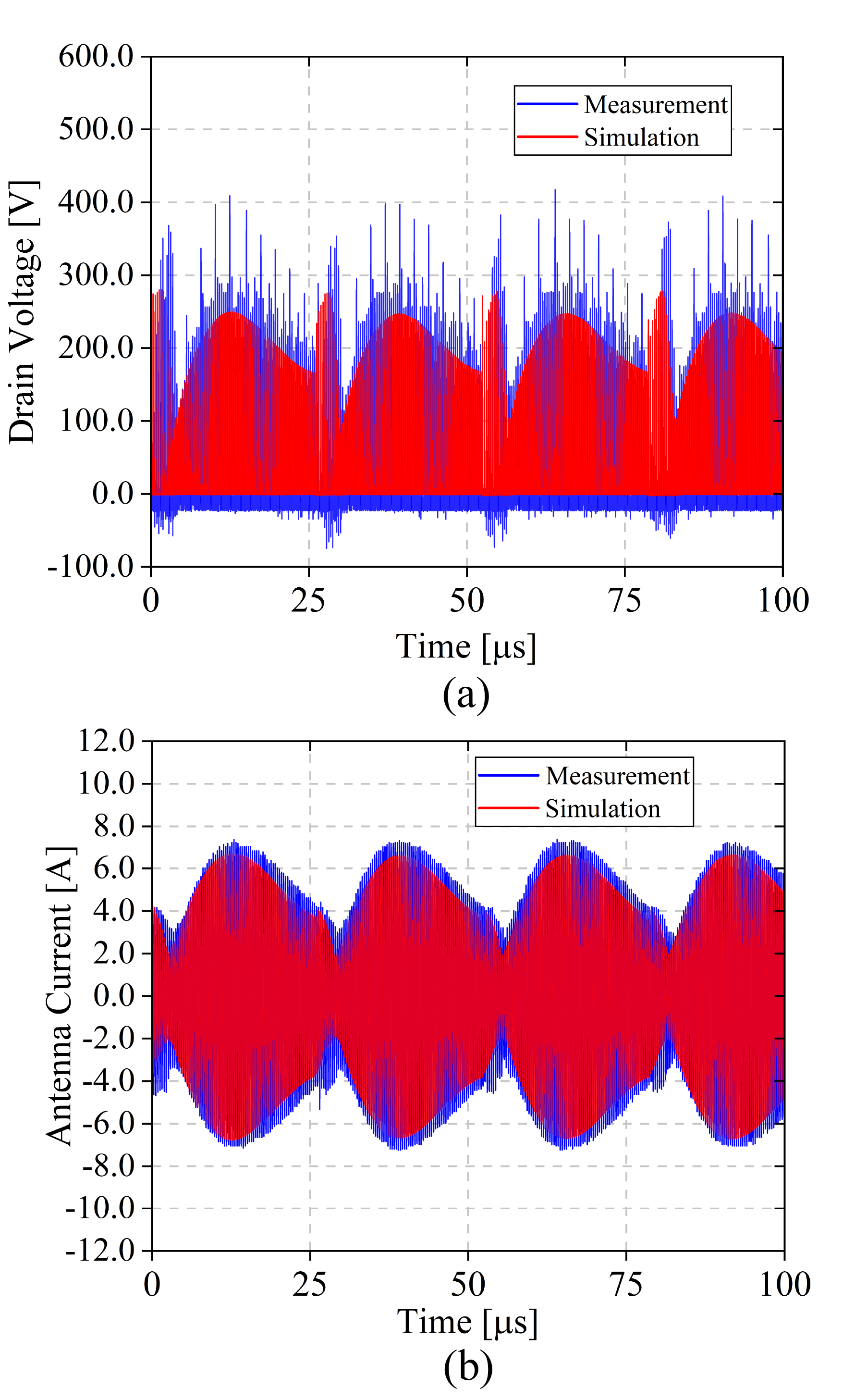}
\caption{Measurement and simulation results for PSK modulated signal with $f$ = 3.07 MHz. $V_{DC}$=40 V, and measured $I_{DC}$ = 1 A. The measured and simulated efficiency is approximately 60$\%$ and 40$\%$ respectively.}
\label{PSKmod}
\end{figure}

We also experimentally demonstrated BPSK modulation with our active antenna, which is less susceptible to noise than ASK modulation. Phase modulation is implemented by directly phase modulating the gate driver signal. During phase changes in the signal, large voltage spikes may abruptly appear across the transistor. This is likely due to the timing at which the transistor turns off relative to the amount of current in the load. The switching of the transistor can cause an abrupt current change in the inductance of the RLC antenna load, which causes a large voltage spike at the drain (see pp. 995-1035 of \cite{PowerE}). These voltage spikes, however, are not present in the antenna current and do not adversely impact the radiated fields from the antenna, which are proportional to the antenna current. Fig. \ref{PSKmod} shows the drain voltage and the antenna current for this experiment. We used a 3.07 MHz BPSK modulated signal at a high bit rate of 38 kb/s which occupies a bandwidth of 38 kHz or more. These simulation and experimental results demonstrate the feasibility of achieving BPSK modulation using the proposed antenna design.

We suspect that the additional drain voltage spikes observed in the measurement compared to the simulation are in part due to nonlinearities not accounted for in the SPICE model used, or due to model inaccuracies in the gate driver. These factors can cause the soft switching conditions of the fabricated prototype to be different from what was simulated, resulting in large voltage spikes (for more information see pp. 995-1035 of \cite{PowerE}). This is also believed to be the cause of the discrepancies observed between the measured and simulated efficiency values (60$\%$ and 40$\%$ respectively). The efficiency is especially sensitive to the transistor’s behavior right at the moment of switching and its accurate prediction requires using models that can precisely account for all the transistor nonlinearities and the transient behavior of the gate driver circuit when the phase of the control circuit is abruptly changed to achieve PSK modulation. Nevertheless, despite these discrepancies, the drain voltage spikes are not present in the antenna current (and thereby the radiated fields), since the abrupt change of antenna current in the inductor is the likely cause for their presence in the drain voltage of the transistor.
%The modulated antenna current is not only phase modulated, but also amplitude modulated. This will affect the BER and the spectrum of the modulated signal. This needs to be taken into account when considering the performance of this modulation technique in comparison to a conventional passive transmitter with a conventional BPSK signal with no amplitude modulation. In Sections \ref{Nonlin_section} and \ref{passive_comp_sec}, we consider both the spectrum and BER when comparing our transmitter to conventional HF  passive transmitting antennas.

{Similar to the case for ASK modulation, we performed coherent I/Q demodulation of the simulated and measured antenna currents shown in Fig. {\ref{PSKmod}} in Matlab. As before, we confirmed that the modulated waveform can be successfully demodulated and the transmitted bits recovered. The simulated transmitter EVM was determined to be -22.8 dB at a bit rate of 38 kb/s without applying any equalization or additional post processing steps.}

\subsection{FSK Modulation}
FSK modulation is of particular interest, because it can provide very robust long-range HF communications. In particular, multi-FSK (MFSK), has the potential of providing very reliable long-range communications via the ionosphere \cite{Q_MFSK}. In MFSK modulation, different carrier frequencies are used. Each frequency functions as a separate symbol and is transmitted one at a time. In this modulation strategy, the number of bits/symbol is large, which allows for a very low symbol rate for a given number of bits/sec. The low symbol rate mitigates fading and multi-path effects associated with long-range communication via the ionosphere \cite{MFSK_robust}.

\begin{figure}[ht]
\includegraphics[width=8.5cm]{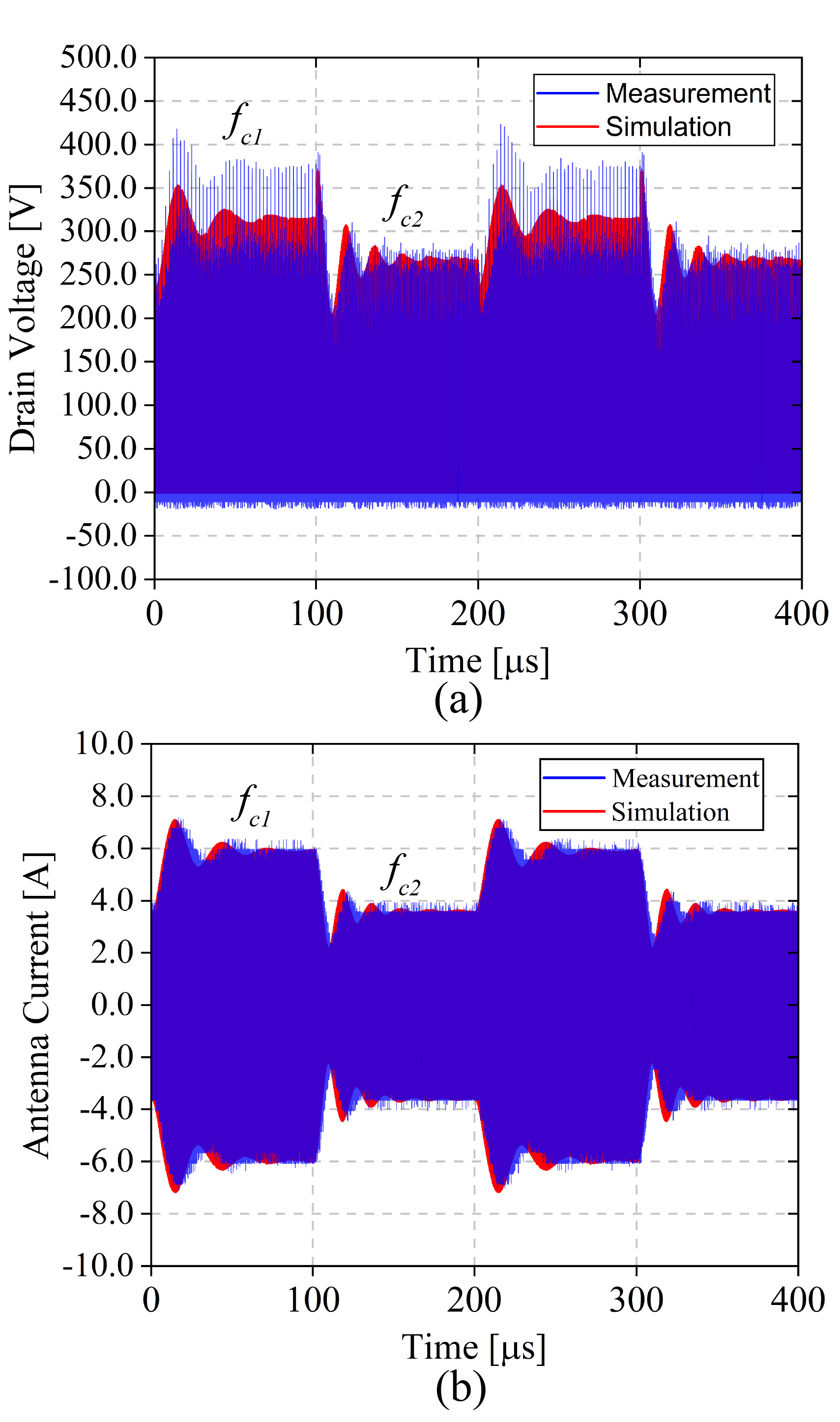}
\caption{Measurement and simulation results for FSK modulated signal with $f_{c1}$ = 3.07 MHz, and $f_{c2}$ = 3.1 MHz. $V_{DC}$=50 V, and measured $I_{DC}$ = 0.5 A. The measured and simulated efficiency is approximately 65$\%$.}
\label{FSKmod}
\end{figure}

Broadband MFSK modulation would involve taking a wideband HF channel of 24 kHz or more and dividing this channel into many carrier frequencies. We demonstrated the ability of our prototype to do BFSK modulation with two carrier frequencies spaced 30 kHz apart (3.07 MHz, and 3.1 MHz). Our signal generator lacked the ability to produce MFSK modulation. However, with a proper signal generator, multiple carrier frequencies for MFSK modulation can be used within this 30 kHz band. Fig. \ref{FSKmod} shows our measurement and simulation results. A 5 kHz key frequency was used in this measurement so that the two different carriers can be easily seen in our measurements, however, this can be easily increased or decreased based on the application. As can be seen, our measurement and simulation results agree very well. Some additional voltage spikes in our measurement are observed in the drain voltage of the transistor, for similar reasons as those discussed in Section \ref{PSKmod_sec}. Observe that the signal is also amplitude modulated. The lower carrier frequency (3.07 MHz), has an amplitude that is about 1.7 times the amplitude of the higher carrier (3.1 MHz). This is due to the RLC antenna model's increased impedance at higher frequencies. The decrease in radiated power at higher frequencies can also be seen in Fig. \ref{CWop2}. This means that the bits at 3.1 MHz would be more susceptible to bit errors in the presence of noise than the bits at 3.07 MHz. We take this into consideration when we compare the performance our FSK modulated transmitter in Section \ref{passive_comp_sec} with an equivalent passive design. This was done by using the average energy-per-bit as the reference for comparison. However, a dynamic DC power supply can correct for this amplitude modulation. By increasing the DC supply voltage during the 3.1 MHz carrier, and decreasing the supply voltage during the 3.07 MHz carrier, the amplitude of the two carriers could be adjusted to the same level. This is a promising technique for achieving very high bandwidths with MFSK modulation.

Another advantage of the FSK modulation technique is that there is only a small reduction in efficiency compared to CW operation. Because this modulation technique only involves slight changes in the carrier frequency, the ZVS condition was maintained in our measurements even during transitions between the carrier frequencies. Our measured and simulated efficiency are both approximately 65$\%$. {Similar to the previous cases, we exported the measured and simulated antenna currents for the FSK modulation and post processed them in Matlab to perform demodulation and detection. As in the previous cases, both signals were successfully demodulated and the transmitted bits recovered. Since traditional I/Q demodulation is not performed for detection of FSK signals, EVM was not calculated in this case.}

\subsection{Evaluation of nonlinearity}
\label{Nonlin_section}
In a real-world system, it is critical that the transmitter does not radiate significant out-of-band harmonics. By taking an FFT of the antenna current, we estimated the radiated nonlinearities in the proposed active antenna. Nonlinearities near the frequency of operation, and at the harmonics of the carrier frequency were examined.

\begin{figure}[ht]
\includegraphics[width=8.5cm]{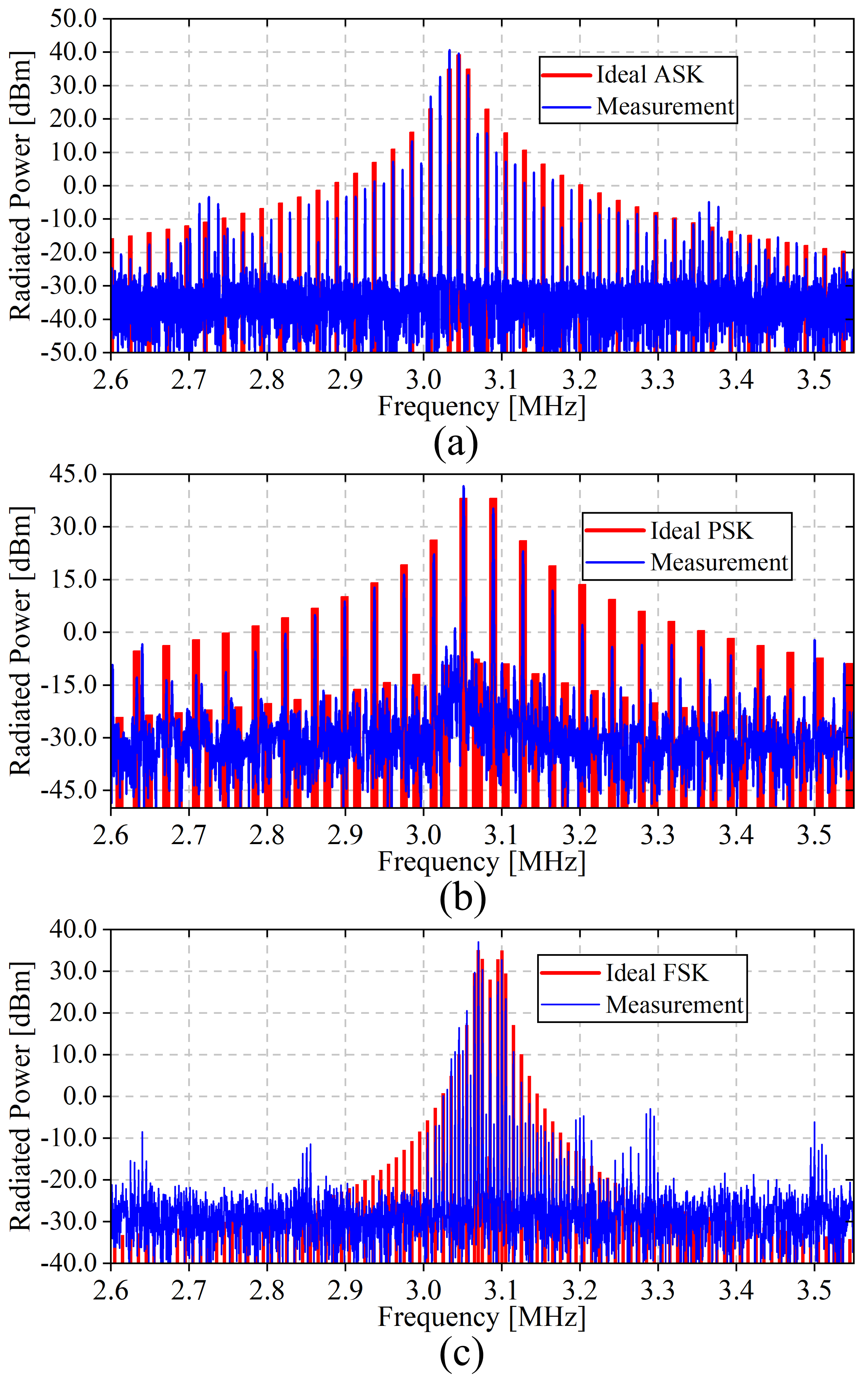}
\caption{Radiated spectrum for each modulation case. (a) ASK, 24 kbits/sec. (b) PSK, 38 kbits/sec. (c) FSK, $f_{c1}$ = 3.07 MHz, $f_{c2}$ = 3.1 MHz, 10 kbits/sec.}
\label{FFT_inband}
\end{figure}

\textbf{Intermodulation Distortion:} Due to nonlinearities, power amplifiers generate what is known as intermodulation distortion, which widens the bandwidth of the signal, and can interfere with neighboring channels \cite{maxim}. Typically, intermodulation distortion is measured by the input of a two-tone signal and measuring the magnitude of the generated third order products. However, the active class-E ESA examined in this paper generates the modulated RF signal right at the antenna level and is not designed to amplify an input RF signal. Thus, the standard two-tone test used in conventional PAs cannot be performed for this non-LTI switch-mode ESA. Instead, to assess the generation of nonlinear signals near the communication band, we measured the spectrum of the antenna current for each modulation case and compared that with that of an ideal, linear modulated signal (at the same frequency and bit rate). We generated the ideal modulated signals with Keysight ADS by generating an ideal ASK, PSK, or FSK signal, and then filtering the signal with a matching network that provided a $-$10 dB match over the bandwidth that contained the signal. The results for each modulation case are shown in Fig. \ref{FFT_inband}. We expressed the spectrum in dBm by computing the amount of radiated power based on the measured current, and the measured 1.43 $\Omega$ radiation resistance of the antenna model.

For ASK modulation, there exists two nonlinear signals generated at approximately 2.72 and 3.37 MHz. The peak of these signals is approximately 7 dB greater than what would be expected from an ideal ASK signal, with no nonlinearity. For PSK modulation, there are no signals near the band of interest that differ significantly from an ideal PSK signal. The case of FSK modulation contains the most significant generation of nonlinear signals near the communication band, namely at 2.62, 2.85, 3.2, 3.3, and 3.5 MHz. However, all of these nonlinear signals are approximately 40 dB or more below the carrier signals.

We point out that the magnitudes of these radiated nonlinear signals relative to the carrier experience little change if the total radiated power is increased. This is in contrast to typical intermodulation distortion in conventional, linear PAs. The third-order product signals in a typical PA will increase by 3 dB for every 1 dB of increase in the carrier \cite{maxim}. However, in our class-E active antenna prototype, the increase in the generated nonlinear signals was significantly less than 3 dB for a 1 dB increase in the carrier power. Specifically, in each modulation case, we varied the supply voltage from the nominal values in Fig. \ref{FFT_inband} to see if there was any decrease or increase in the nonlinearities relative to the carrier. In each case, the radiated nonlinear signals increased approximately 1 dB for every 1 dB of increase in the carrier. This is because the transistor is consistently operated as a switch, no matter what the level of radiated power is. This is in contrast to a conventional commercially available biased class A or AB amplifier or an active antenna operating in such a class such as the one reported in \cite{Class_AB_ESA}, where the nonlinearities will increase when the input power is increased, due to the transistor approaching saturation at higher power levels.

\begin{figure}[ht]
\includegraphics[width=8.5cm]{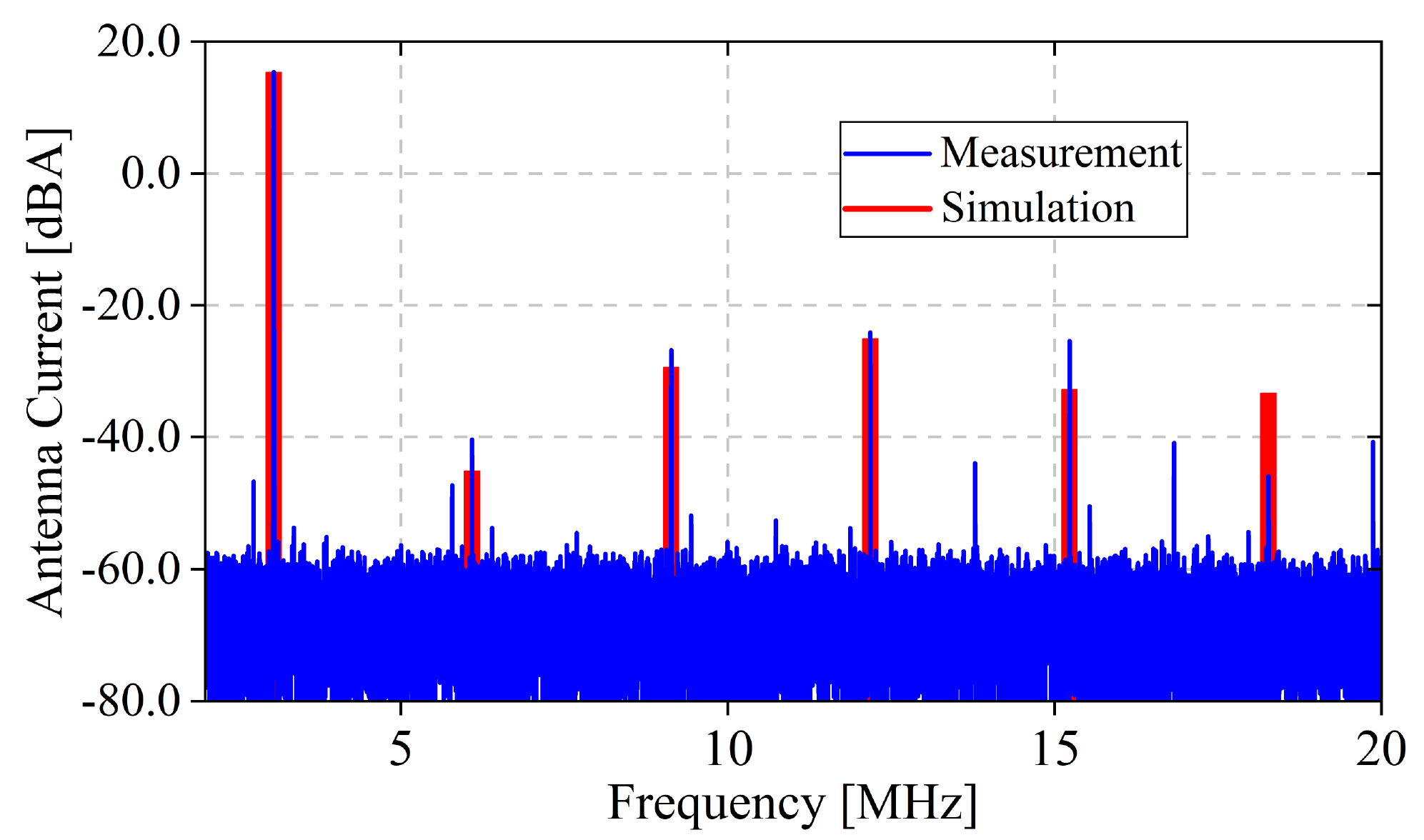}
\caption{Simulated and measured antenna current broadband harmonics for CW operation at 3.045 MHz.}
\label{FFT_outband}
\end{figure}

\textbf{Radiated Harmonics:} We also evaluated the higher-order radiated harmonics. First, we measured the broadband spectral content of the current through the antenna model with a CW signal. Our measurement results are shown in Fig. \ref{FFT_outband}. The higher order harmonics were also simulated with a harmonic balance simulation in Keysight ADS. The difference between simulation and measurement is at most 10 dB at the $6^{th}$ harmonic (18 MHz), but on average agrees more closely within 5 dB.

As we are using a lumped element antenna model, it is not possible to precisely measure the level of radiated harmonics based on the harmonics of the current. This is the case since the effective radiation resistance of the antenna model is different at the harmonic frequency values than a realistic antenna. However, from our measurements we determined a conservative estimate for the radiated power levels by estimating the radiation resistance at each harmonic frequency. Our estimation was based on the fact that the radiation resistance of a dipole or monopole-type ESA increases with the square of the frequency \cite{balanis_2016}. For example, we estimate that the radiation resistance of the antenna at the 2$^{nd}$ harmonic (approximately 6 MHz), would be $1.43\times 2^{2} = 5.72$ $\Omega$. Using these radiation resistance values, we estimated the harmonic radiated power levels and show our results in Table \ref{RadHarmTab}. We estimated the levels of the radiated power for CW, as well as every modulation case. The worst-case scenario is with PSK modulation, where the harmonic level at 12 MHz is -22.5 dBc. Note that these harmonic levels can also be decreased in principle by increasing the impedance of the RLC antenna load at higher frequencies.

\begin{table}[!ht]
\caption{Estimated radiated broadband harmonics for different modulation cases.}
\label{RadHarmTab}
\centering
\begin{tabular}{|l|l|l|l|l|}
\hline
 Freq.[MHz]& CW [dBc] & ASK [dBc] & PSK [dBc] & FSK [dBc] \\
\hline
$\approx$ 6 & -49.8 & -38.8 & -40.4 & -48.5 \\
\hline
$\approx$ 9 & -32.7 & -27.9 & -28.2 & -34.7 \\
\hline
$\approx$ 12 & -27.5 & -24.4 & -22.5 & -29.9 \\
\hline
$\approx$ 15 & -26.9 & -30.8 & -29.3 & -29.9 \\
\hline
\end{tabular}
\end{table}

\section{Comparison with Passive Matching}
\label{passive_comp_sec}
\begin{figure}[ht]
\includegraphics[width=8.9cm]{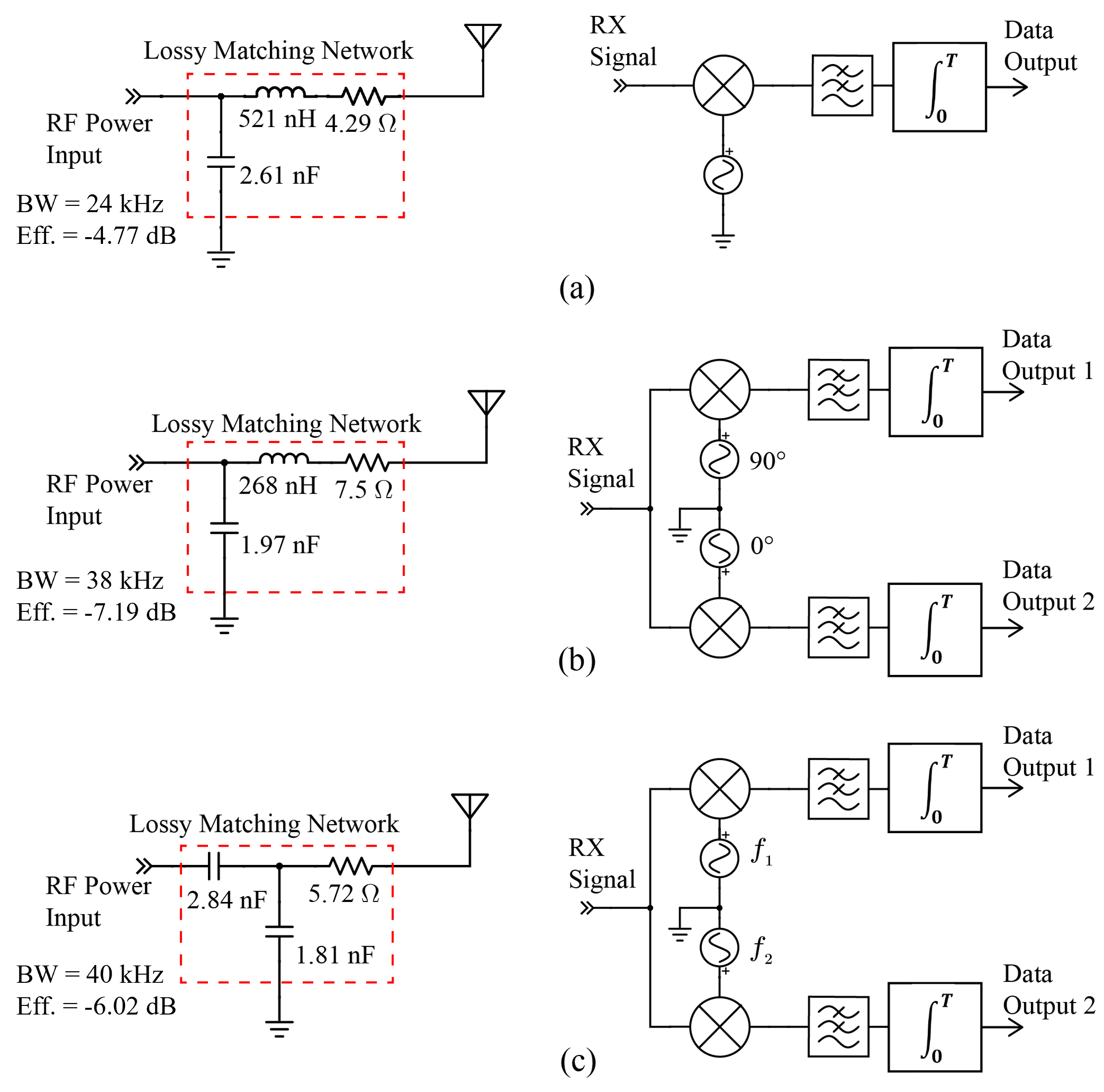}
\caption{Passive transmitter designs for comparison with the demonstrated active antenna. The corresponding receiver is shown in each modulation case. (a) ASK modulation, (b) PSK modulation, (c) FSK modulation.}
\label{Comparison}
\end{figure}

We conducted a thorough analysis comparing the performance of our active antenna with conventional passively matched HF antennas. Specifically, we designed and simulated conventional passive matching networks for each of the three modulation cases considered in this work (ASK, PSK, and FSK). Each passive design assumes the same lumped antenna model used in our experiments with a 1.43 $\Omega$ radiation resistance and negligible loss resistance. This was to ensure a fair comparison. Because the lumped element antenna model only has a -10 dB bandwidth potential of 6 kHz, we had to design a lossy impedance matching network for each case to accommodate the required signal bandwidth. We added just enough loss to each matching network to allow for the necessary bandwidth, but would also not incur unnecessary losses. To determine an appropriate amount of loss to add to the antenna, we first determined the required bandwidth for each modulation case. For BASK and BPSK modulation, most of the signal power is contained within the frequencies between $f_{c}-R_{b}/2$ and $f_{c}+R_{b}/2$ (where $f_{c}$ is the carrier frequency, and $R_{b}$ is the bit rate in bits/sec.) \cite{commtextbook}. Similarly, for BFSK modulation, most of the signal power is contained within a bandwidth of $\Delta f+R_{b}$, where $\Delta f$ is the spacing between the two carrier frequencies, and $R_{b}$ is the bit rate in bits/sec. Therefore, for each of our modulation cases, the required bandwidth for each passive design would be 24 kHz, 38 kHz, and 40 kHz for ASK, PSK, and FSK respectively. We used Optenni Lab to design lossy impedance matching networks to achieve these required bandwidths. Only the amount of loss needed to achieve the necessary bandwidth was used. The matching networks were designed to provide at least an approximately -10 dB reflection coefficient across the band to avoid damage to the PA, and to not attenuate the sidebands of the modulated signal. Fig. \ref{Comparison} shows the lossy matching networks designed for each modulation case.

After synthesizing each passive matching network design, we simulated each modulation case for the passively-matched antenna in Keysight ADS with a time-domain solver. Each passive case was designed to achieve the same {bit error rate} in an additive white gaussian noise channel (AWGN) as our active design. The amount of total input power (RF+DC) required to achieve the same BER was used to compare our active antenna with a conventional passive antenna. To determine the amount of input power required to achieve the same {BER} as our active antenna, we implemented an ideal, synchronous demodulator (using a multiplier and integrate-and-dump architecture \cite{commtextbook}) in Keysight ADS in both our active and passive designs. We adjusted the radiated power of each passive design so that the integral of the integrate-and-dump demodulator was the same as that obtained for our active antenna. If the integration values are the same, this means that both systems have the same BER in the presence of an AWGN channel. Additionally, the performance characteristics of each passive ESA design is reported in terms of bandwidth and radiation efficiency ($P_{rad}/P_{in}$). Note that the reported efficiency is assuming that the 1.43 $\Omega$ resistance of the antenna model is purely radiation resistance. However, this assumption is irrelevant for the sake of comparing performance with the active class-E antenna since the same assumption is used in evaluating the performances of both active and passive matching cases.

In addition to losses that must be added to a passive matching network to achieve the required bandwidth, it is also important to consider losses in a conventional PA. Typical commercially available PAs used at the HF band are class A, or AB. A class-A amplifier can have a maximum theoretical DC-to-RF efficiency of $e_{max}=0.5/\text{PAPR}$, where PAPR is the peak-to-average power ratio \cite{PAPR}. In our comparison, we assumed the maximum efficiency of a class-A amplifier (including the PAPR associated with each modulated signal), for the efficiency of the PA. However, it should be noted that a real-world amplifier will most likely have less efficiency. We summarize the results of our comparison in Table \ref{PassComp}. In all three cases, we show significant bandwidth-efficiency product improvement. Specifically, the bandwidth-efficiency product is improved by 5.4 dB, 8.7 dB, and 9.8 dB respectively for BPSK, BFSK, and BASK. {The absolute efficiency of the class-E antenna reported in this work may be reduced if a radiating element with a lower radiation efficiency is used or if the antenna is placed close to lossy ground (e.g., placing the antenna shown in Fig. {\ref{Concept}}(b) in parallel with earth). This reduction, however, impacts both the proposed class-E active antenna and its passively-matched counterparts. Therefore, the bandwidth-efficiency product improvement factors shown in Table {\ref{PassComp}} are not expected to change significantly. In applications where monopole versions of such an antenna are mounted on ships and naval vessels, the presence of earth is not expected to reduce the antenna efficiency since seawater is an excellent conductor at HF frequencies.}

%Note that even if a PA with a theoretical DC to RF efficiency of 100$\%$ was used in conjunction with the passively-matched ESAs, significant $\beta \times \eta$ improvements of 3.8 dB, 2.44 dB, and 5.7 dB respectively for BASK, BPSK, and BFSK modulations would still result.

\begin{table}[!ht]
\caption{Comparison with Conventional HF Antennas. }
\label{PassComp}
\centering
\begin{tabular}{|l|l|l|l|}
\hline
 Modulation Type & $P_{in,a}$ [W] & $P_{in,p}$ [W] & $(\beta \times \eta)_{imp}$  [dB] \\
\hline
BASK & 40 & 380 & 9.8 \\
\hline
BPSK & 40 & 140 & 5.44 \\
\hline
BFSK & 25 & 186 & 8.7 \\
\hline
\end{tabular}
\flushleft
*$P_{in,a}$ and $P_{in,p}$ is the total input power (RF+DC), for each active and passive case, respectively. Since bandwidth ($\beta$) is the same for both passive and active ESAs, bandwidth-efficiency product improvement   is $(\beta \times \eta)_{imp}$ = 10$\log(P_{in,p}/P_{in,a}).$
\end{table}

Besides considerable improvement of $\beta \times \eta$, there are several additional practical considerations which make our active design more favorable than a conventional passive antenna. First, the operation of the active antenna is relatively immune to changes in the antenna impedance due to changes in environment. This is because the operational bandwidth of our active antenna is so wide, as shown in Fig. \ref{Concept}(c). As long as the sub-optimum operating conditions are satisfied \cite{SubOpt_analysis}, the operation of the circuit will be affected very little. {Additionally, the proposed class-E active antenna can be used with a wide range of dipole or monopole-type radiating elements. The only condition required is that the antenna is either self resonant (e.g., meander type miniaturized dipoles) or is made resonant with the aid of an external inductor (e.g., the case presented in this work).}

Our active design also offers another important advantage in situations when the full power capability of the amplifier is not necessary to communicate over a given distance. When a conventional PA is operated below its rated power, its efficiency inevitably drops. However, this is not the case with the reported active class-E ESA. The radiated power is proportional to the square of the DC supply voltage, and the efficiency of the active antenna is independent of the radiated power level. If a lower power level is desired, all that is necessary is to adjust the DC supply voltage to a lower level, without any compromise in efficiency.

\section{Discussion}

\begin{table*}[!ht]
\caption{Comparison of the presented design with other state-of-the-art active or non-LTI electrically-small transmit antennas.}
\label{State-of-the-art}
\centering
\begin{tabular}{|>{\color{black}}p{0.4in}|>{\color{black}}p{1.5in}|>{\color{black}}p{0.74in}|>{\color{black}}p{0.74in}|>{\color{black}}p{0.74in}|>{\color{black}}p{0.5in}|>{\color{black}}p{0.8in}|>{\color{black}}p{0.8in}|}
  \hline
  Reference & Technique & Ant. Impedance & $(k\times a)^*$ & Fractional BW$\times$Efficiency ($\beta\times\eta$) & $P_{rad}$ & Mod. Types \\ \hline
  \cite{Sussman1} & Non-Foster (Class-C) & $1-j224~\Omega$ & 0.28  & 1.18$\%$ & 1.3 W & Unrestricted \\ \hline
  \cite{Tingyen} & Non-Foster + Buffer & $5-j500~\Omega$ & 0.17  &  6.4$\%$ & 40 mW & Unrestricted \\ \hline
  \cite{Class_AB_ESA} & Antenna + Linear Amplifier (Class-AB)& $1.5-j3500~\Omega$ & 0.06  &  0.11$\%$ & 108 mW & Unrestricted  \\ \hline
  \cite{Transitent_state_ant} & Direct Antenna Modulation & $0.1+j14~\Omega ^{**}$  & 0.05  &  $0.18\%$ & 0.5 mW$^{\ast}$ & DSB-AM \\ \hline
  \cite{HEVHFESABFSKDAM} & Direct Antenna Modulation & 0.34 $\Omega$ & 0.10  &  $0.13\%$ & 10 W & BFSK \\ \hline
  This Work & Antenna + Switch-Mode Amplifier (Class-E) & $1.4-j531~\Omega$ & 0.07  &  2.89$\%$ & 64 W & BASK, BPSK, $\&$ BFSK \\
  \hline
\end{tabular}
\flushleft
*~$ka=\frac{2\pi a}{\lambda_0}$ is the wave number and $a$ is the radius of the smallest circumscribing sphere. The value is reported at the lowest frequency of operation of the antenna.\\
**~Exact value not reported in the paper. The values are estimated from other data presented in the reference.
\end{table*}

In this work, we demonstrated a high-power, switch-mode electrically-small antenna operating in sub-optimum class-E mode that offers between 5.4 dB and 9.8 dB improvement in bandwidth-efficiency product compared to a conventional passively matched ESA with the same bit rate, and BER. Table \ref{State-of-the-art} shows a comparison between the switch-mode active ESA reported in this paper and a few other electrically-small transmit antennas reported in the literature that use competing techniques for enhancing the bandwidth-efficiency product of ESA compared to conventional passive matching techniques. As discussed in  Section \ref{sec:introduction}, the input resistances of electrically-small antennas is generally extremely small. This is in sharp contrast to electrically-large resonant antennas where the input impedance is usually purely resistive with input resistance values often close to 50  $\Omega$. Therefore, design of active electrically-small antennas is often far more challenging than designing active antennas where the radiating element is electrically large. As a result, the scope of the comparison presented in Table \ref{State-of-the-art} is limited to electrically-small antennas that use active matching or other competing techniques for enhancing the bandwidth-efficiency product of a transmitting electrically small antenna. Since the antennas reported in Table \ref{State-of-the-art} have different antenna impedances and quality factors and they work at different frequency bands, it is somewhat difficult to perform a one-to-one comparison between them. Nevertheless, it can be observed that the bandwidth-efficiency product\footnote{Bandwidth-efficiency product numbers reported in Table\ref{State-of-the-art} are obtained by calculating the fractional bandwidth of the antenna (in percent) and multiplying that by the reported efficiency value (in linear scale).} of the class-E, active ESA presented in this paper compares very well with other antennas with similar electrical dimensions (i.e., similar $ka=\frac{2\pi}{\lambda}a$ value). Only one antenna reported in Table \ref{State-of-the-art} has a higher $\beta \times \eta$ value \cite{Tingyen}, which is in part due to its larger electrical dimension ($2.42\times$) and lower quality factor. More importantly, however, the effective radiated power level of the class-E switch-mode ESA presented in this paper is significantly higher than those of the other electrically small antenna designs reported in Table \ref{State-of-the-art}. Specifically, a radiated power level of 64 W was measured, with very large peak-to-peak voltage and current swings of 10 kV and 19 A respectively at the antenna terminals. This is significantly greater than what has been previously reported in the literature. Compared with the active ESA reported in \cite{Class_AB_ESA} (see Table \ref{State-of-the-art}), the proposed switch-mode ESA has $26\times$ higher bandwidth-efficiency product and $593\times$ higher radiated power level.

In addition to demonstrating enhancements in the radiated power level and bandwidth-efficiency product, we demonstrated that binary ASK, PSK, and FSK digital modulation schemes can be successfully generated using the proposed active ESA and examined the generation of undesirable nonlinear signals and radiated harmonics. The generated nonlinear signals near the band of communication were all 40 dB or more below the carrier for each modulation case. For out-of-band harmonics, the worst-case-scenario was estimated to be 22.5 dB below the carrier at 12 MHz for PSK modulation. {Additionally, the error vector magnitude of the transmitted ASK and PSK signals were estimated to be -24 dB and -22.8 dB, respectively for bit rates of 24 kb/s and 38 kb/s.} While much remains to be done in evaluating the transient modulation behavior of class-E active ESAs, our work shows very promising results for producing reliable wideband active transmit ESAs in the HF band.

While we demonstrated that the proposed class-E active ESA can generate binary ASK, PSK, and FSK modulations, achieving higher-order modulations such as 16QAM is also of interest in many communications systems that use electrically small antennas. We anticipate that the proposed class-E active ESA may be modified to generate these more complex modulation schemes. For example, to generate a modulated waveform with 16QAM modulation, both the phase and the amplitude of the antenna current must be modified simultaneously. In principle, this can be accomplished by extending the techniques used to generate ASK and PSK modulations in this work to generate more than two different amplitude or phase states. For example, a more complex amplitude control circuit capable of generating 3 different amplitude levels can be combined with a more complex gate driver circuit capable of generating 12 different phase values. This way, by changing both the amplitude and phase of the antenna current at every symbol, each of the 16 different symbols of a 16QAM modulation may be generated. While relatively easy to conceptualize, implementation of such higher-order modulation schemes requires significant additional work that is beyond the scope of the current manuscript and will be left for a future work.

\section*{Acknowledgment}
The authors would like to thank Maggie Muldowney for her assistance with taking photos of our experimental setup.

\end{document}